\documentclass[11pt]{article}
\usepackage{graphicx} 
\usepackage{amsmath, amssymb, mathtools}
\usepackage{amsthm} 
\newtheorem{theorem}{Theorem}

\newtheorem{remark}{Remark}
\newtheorem{hypothesis}{Hypothesis}

\newtheorem{corollary}{Corollary}
\newtheorem{lemma}{Lemma}
\usepackage{hyperref}
\usepackage[nameinlink]{cleveref}
\usepackage[style=alphabetic, backend=biber, maxbibnames=99, maxalphanames=4, minalphanames=3]{biblatex} 
\newcommand{\calB}{\mathcal{B}}
\newcommand{\calQ}{\mathcal{Q}}
\newcommand{\calP}{\mathcal{P}}

\usepackage{fullpage}

\usepackage{xspace}

\NewDocumentCommand{\MaxkSat}{O{k}}{%
  \textnormal{\textsf{Max}-}\ensuremath{#1}\textnormal{\textsf{SAT}}\xspace
}

\def \eps{\varepsilon}
\DeclarePairedDelimiter{\floor}{\lfloor}{\rfloor}
\DeclarePairedDelimiter{\ip}{\langle}{\rangle}
\DeclareMathOperator{\poly}{poly}
\usepackage{xcolor}
\usepackage{tikz}
\usetikzlibrary{arrows.meta, backgrounds}

\DeclareMathOperator{\Ckt}{C}
\DeclareMathOperator{\Rk}{R}

\DeclareMathOperator{\PR}{PR}
\DeclareMathOperator{\Span}{span}

\newcommand{\sfT}{\mathsf T}
\newcommand{\row}{\mathrm{row}}

\renewcommand{\setminus}{\smallsetminus}
\renewcommand{\emptyset}{\varnothing}

\NewDocumentEnvironment{introtheorem}{m +b}
  {%
    \expandafter\gdef\csname storedtheorem@#1\endcsname{#2}%
    \begin{theorem}\label{#1}%
      #2%
    \end{theorem}%
  }
  {}

\newcommand{\reminderheading}{}
\theoremstyle{plain}
\newtheorem*{reminderinner}{\reminderheading}

\NewDocumentEnvironment{reminder}{m}
  {%
    \renewcommand{\reminderheading}{Reminder of \Cref{#1}}%
    \begin{reminderinner}%
      \ifcsname storedtheorem@#1\endcsname
        \csname storedtheorem@#1\endcsname
      \else
        \PackageError{theorem-reminders}
          {No stored theorem with label '#1'}
          {Use introtheorem with this label before its reminder.}%
      \fi
  }
  {\end{reminderinner}}

\newcommand{\PIP}{\mathrm{PIP}}

\title{Finding Hypercliques Faster Than Brute Force}

\author{Josh Alman\thanks{Columbia University. Supported in part by NSF Grant CCF-2238221 and a Packard
Foundation Fellowship.} \and Baitian Li\thanks{Columbia University. Supported in part by NSF Grant CCF-2238221 and a Packard
Foundation Fellowship.} \and Neha Pant\thanks{Columbia University. Supported in part by a Columbia Engineering Grossman Fellowship.} \and Kevin Pratt\thanks{Columbia University. Supported in part by NSF Grant CCF-2238221 and a Packard
Foundation Fellowship.} \and Ryan Williams\thanks{MIT. Work supported in part by NSF CCF-2420092 and the AI Security Institute (AISI).}}
\date{}

\begin{document}

\maketitle
\begin{abstract}
We show that for any fixed $k$, there exists $\varepsilon_k > 0$ such that, given a $(k-1)$-uniform hypergraph $H$ on $n$ vertices, one can count the number of $k$-hypercliques in $H$ in $O(n^{k - \varepsilon_k})$ time. This refutes the hyperclique conjecture of Lincoln, Vassilevska Williams, and Williams (SODA 2018). Our result follows from the uniform construction of a family of arithmetic circuits of size $O(n^{k - \varepsilon_k})$ for the ``$k$-hyperclique tensor", a natural higher-order generalization of the matrix multiplication tensor.

As corollaries, it follows that 
\begin{itemize}
    \item For every $k$, there is an $\eps > 0$ such that \MaxkSat{} can be solved in $2^{(1-\eps)n} \cdot \poly(m,n)$ time.
    \item For every $k$, there is an $\eps > 0$ so that arithmetic progressions of length $k$ in subsets of $\{0,1,\ldots,n\}$ can be detected in $n^{2-\eps}$ time. 
\end{itemize}
Both results refute other prominent conjectures in the literature.

\vspace{0.5cm}
\textit{The main result was obtained by ChatGPT 6 Pro. The full chat is available at }
\begin{center}
\url{https://chatgpt.com/share/6acf869d-a8f0-83ea-a00d-35492f5a3af5}
\end{center}
\end{abstract}

\thispagestyle{empty}
\clearpage
\pagenumbering{arabic}

\section{Introduction}

One of the most striking algorithms avoiding exhaustive search is the $n^{\omega+o(1)} \leq n^{2.25+o(1)}$~\cite{OpenAI2026MatrixMultiplication} time algorithm for detecting (and finding) a triangle in an $n$-node graph, first reported by Itai and Rodeh~\cite{IR78}. Taking $A$ to be the $n \times n$ adjacency matrix, $\operatorname{tr}(A^3)$ is exactly six times the number of triangles, so computing it with fast matrix multiplication decisively beats the naive $O(n^3)$ time of exhaustive search. (By recursing on subgraphs, we can also \emph{find} triangles in analogous running time.) Since Itai and Rodeh, \emph{many} other improvements in graph algorithms have been powered by fast matrix multiplication, such as transitive closure, shortest paths, matching, cycle and subgraph detection, and many others; see the surveys and lecture notes of Zwick~\cite{Zwick15MatrixGraphAlgorithms} and Vassilevska Williams~\cite{VassilevskaWilliams18ICM,VassilevskaWilliams21MatrixGraphAlgorithms}.

It is natural to ask whether similar algorithmic improvements are possible for problems on generalizations of graphs, such as $3$-uniform hypergraphs whose edge sets are $3$-sets of vertices, or more generally $k$-uniform hypergraphs for constant $k \geq 3$. A starting problem, generalizing triangle finding, is that of finding a \emph{$k$-hyperclique in a $(k-1)$-uniform hypergraph} for $k \geq 3$. Williams~\cite{Wil05} raised the problem for $k=4$, noting that a faster $4$-hyperclique algorithm could be applied to solve $\MaxkSat[3]$ faster than exhaustive search. This reduction was generalized for arbitrary $k$ by \cite{lincoln2018tight}: faster $(k+1)$-hyperclique finding in $k$-uniform hypergraphs would yield a faster algorithm for $\MaxkSat$. They proposed a hardness hypothesis for $k$-hyperclique finding:

\begin{hypothesis}[$k$-Hyperclique Hypothesis] \label{hyp:k-hc} For every $k \geq 4$, there is no $\eps > 0$ such that $k$-hypercliques in $(k-1)$-uniform hypergraphs can be detected in $O(n^{k-\eps})$ time.
\end{hypothesis}

As evidence for their hypothesis, they proved that \emph{unlike} matrix multiplication which famously has subcubic tensor rank (starting with Strassen~\cite{Str69Gaussian}), the tensor rank of the natural $(k-1)$-dimensional generalization of matrix multiplication (precisely the kind of task one needs to solve to compute $k$-hypercliques, see \Cref{sec:hc} for a definition) is $\Theta(n^k)$. This led \cite{lincoln2018tight} to conclude that radically different algorithmic techniques would be needed to refute the $k$-Hyperclique Hypothesis, if it is possible at all. Further work extensively studied consequences of proving and refuting the hypothesis~\cite{DBLP:conf/icalp/MathialaganWX22,DBLP:conf/icalp/BringmannCFK22,DBLP:conf/focs/DalirrooyfardJW22,DBLP:conf/soda/Chan23,DBLP:conf/latin/AbboudFS24,DBLP:conf/innovations/GokajK25,DBLP:conf/icalp/FischerKRS25,dudek2026classifying,brand2026partition,brand2026beyond}. Notably, \cite{dudek2026classifying} recently showed that if the $k$-Hyperclique Hypothesis were false, then one could detect arithmetic progressions of length $k$ in an input set $A \subseteq \{0,1,2, \ldots,n\}$ in $O(n^{2-\eps})$ time for some $\eps > 0$.

By the work of \cite{Wil05,lincoln2018tight} mentioned above, the $k$-Hyperclique Hypothesis was also believed to be true because it followed from another algorithmic hypothesis about the complexity of \MaxkSat:

\begin{hypothesis}[Max-$k$-SAT Hypothesis] \label{hyp:maxksat} For every $k \geq 3$, there is no $\eps > 0$ such that $\MaxkSat$ on $n$ variables can be solved in $O(2^{(1-\eps)n})$ time.
\end{hypothesis}

This hypothesis had a vast literature as well, with a body of work studying algorithms and hardness for \MaxkSat~\cite{DBLP:conf/sat/Williams03,DBLP:conf/sat/DantsinW06,DBLP:conf/csr/KulikovK07,DBLP:conf/focs/ImpagliazzoPS13,DBLP:conf/sat/SakaiST14,DBLP:conf/iwpec/SakaiSTT15,DBLP:conf/sat/ChenS15,DBLP:conf/focs/AlmanCW16,DBLP:conf/stoc/AbboudBDN18,DBLP:conf/focs/BennettGS17,DBLP:conf/coco/BringmannFK19,DBLP:conf/innovations/AlmanW20,DBLP:conf/esa/Jansen020,DBLP:conf/icalp/BringmannCFK22,DBLP:conf/soda/BelovaKMRRS24,DBLP:conf/esa/Spalding-Jamieson25,DBLP:journals/corr/abs-2507-22293,DBLP:conf/stacs/ChukhinKMS26}. There was also a ``sparse'' version of the \MaxkSat hypothesis which asserted that $O(n)$-clause instances were already hard~\cite{DBLP:conf/innovations/AlmanW20}. 

\paragraph{Our Results.} We prove that the $k$-Hyperclique Hypothesis is false. In fact, we show that for every $k \geq 3$, there is an $\eps > 0$ such that the appropriate $k$-dimensional generalization of matrix multiplication (a.k.a., the $k$-wise matrix product~\cite[Section~5.1]{BS21}) can be computed on Boolean matrices in $O(n^{k+1-\eps})$ time. As an immediate application, this allows us to \emph{count} the number of hypercliques, so that for every $k$, even $\#\MaxkSat$ can be solved in $2^{(1-\eps_k)n}$ time for some $\eps_k > 0$. 

Our main technical result is that the natural polynomial for enumerating hypercliques (a sum over all possible hypercliques which takes the product of all hyperedges in the hyperclique, see \Cref{sec:hc} for a definition) has uniform arithmetic circuits that are significantly smaller than $n^k$ size.

\begin{introtheorem}{thm:hyperclique}
For every fixed $k\ge 3$, the $k$-hyperclique polynomial $\Delta^{(k)}_n$ has uniformly constructible circuits of size $O(n^{k-\eps_k})$, where $\eps_k = 2^{-\Theta(2^k)}$, with all constants in ${\mathbb Z}[1/2]$. In particular, there is an algorithm that, given $(1^n, 1^k)$ as input, outputs a circuit of $O(n^{k-\eps_k})$ size in $n^{k-\eps_k+o(1)}$ time.     
\end{introtheorem}

Many interesting consequences follow. First of all, we can easily count the number of hypercliques in a hypergraph, by evaluating such a polynomial.

\begin{introtheorem}{thm:countHC}
For every fixed $k\ge 3$, there is an algorithm that, given any
$(k-1)$-uniform hypergraph $H$ on $n$ vertices, outputs the number
of $k$-hypercliques in $H$ in $n^{k-\eps_k+o(1)}$ time, where
$\eps_k$ is the constant in \Cref{thm:hyperclique}.
\end{introtheorem}

Furthermore, applying the Baur--Strassen Theorem~(\Cref{lem:Baur--Strassen}), we can compute the natural $k$-dimensional generalization of matrix multiplication faster (i.e., the $k$-wise matrix product), as well.

\begin{introtheorem}{thm:k-dim-prod}
For every $k\ge 3$, there is an $\eps>0$ such that
$\mathrm{MP}_k(A_1,\dots,A_k)$ can be computed on
$A_1,\ldots,A_k$, each with $n^k$ Boolean entries, in
$O(n^{k+1-\eps})$ time.
\end{introtheorem}

Applying these results, we can refute a much weaker hypothesis than just the \MaxkSat Hypothesis (Hypothesis~\ref{hyp:maxksat}). In fact, we solve a much more general {\sf Max-Degree-$k$CSP} problem~\cite{lincoln2018tight}. A Boolean function on $n$ variables is a \emph{degree-$k$ constraint} if it has degree at most $k$ in its unique multilinear polynomial representation over the $\{0,1\}$ basis. (Note that \MaxkSat represents the very special case where each polynomial is over at most $k$ variables.) The {\sf Max-Degree-$k$CSP} problem is: given any set of degree-$k$ constraints over $n$ Boolean variables, find a Boolean assignment that sets a maximum number of constraints to true.

\begin{introtheorem}{thm:maxksat}
For every $k\ge 3$, there is a $\delta_k>0$ such that
{\sf Max-Degree-$k$CSP} with $n$ variables and any number of
$m$ constraints can be solved in
$2^{(1-\delta_k)n}\poly(n,m)$ time.
\end{introtheorem}

Our results also refute a recently proposed hypothesis on the difficulty of finding short arithmetic progressions in a set of numbers. The $k$-AP Hypothesis~\cite[Hypothesis~1.3]{dudek2026classifying} (see also~\cite{fischer2026beyond}) states that for every $k\ge 4$ there is no $\varepsilon>0$ such that one can determine whether a subset of $\{0,1,\dots,n\}$ contains a nontrivial arithmetic progression of length $k$, in $O(n^{2-\varepsilon})$ time. (Note that for $k=3$ the problem can be solved in near-linear time by FFT, so the hypothesis starts at $k=4$.) We give an improved algorithm:

\begin{introtheorem}{thm:kap}
For every fixed $k\ge 4$, we can decide whether a set
$A\subseteq\{0,\dots,n\}$ contains a nontrivial $k$-term
arithmetic progression in
$O(n^{2-2\varepsilon_k/k})$ time.
\end{introtheorem}

Finally, our results also have bearing on a well-known problem in database theory. The \emph{$k$-dimensional Loomis--Whitney join} $\mathrm{LM}_k$ has $k$ attributes and one relation on each $(k-1)$-subset of the attributes. Its \emph{Boolean} version is exactly $k$-hyperclique on a $(k-1)$-uniform hypergraph, and its \emph{counting} version is the polynomial $\Delta^{(k)}_n$ evaluated on the relations. The AGM bound~\cite{AGM13} shows that the output of  $\mathrm{LM}_k$ on relations of total size $m$ has size at most $m^{k/(k-1)}$, and worst-case optimal join algorithms~\cite{NPRR18} can list all answers in $\tilde{O}(m^{k/(k-1)})$ time. The question of whether the Boolean and counting versions can be evaluated faster has been a prominent open problem in database theory; see e.g., the PODS 2025 tutorial of Mengel~\cite[Theorem~3.5]{Mengel25}, who shows that the $k$-Hyperclique Hypothesis implies no $O(m^{k/(k-1)-\eps})$ time algorithm exists. \Cref{thm:countHC} directly gives a faster algorithm for dense instances. By a standard high degree/low degree trick just as in \cite{AYZ97}, we can also obtain that for every $k$ there is an $\eps > 0$ and an $O(m^{k/(k-1)-\eps})$ time algorithm for these joins, for all sizes $m$.\footnote{In graph-theoretic language, say a vertex has low degree if it appears in at most $D$ hyperedges. Using standard algorithms (\cite{NPRR18}) we can count all hypercliques containing a low-degree vertex in $\tilde{O}(m \cdot D^{1/(k-2)})$ time. There are at most $(k-1)m/D$ vertices without low degree. By our results, we can count the hypercliques over just those vertices in time $O((km/D)^{k-\eps})$, for some $\eps > 0$. Setting $D$ to balance the two, we obtain an $O(m^{k/(k-1) - \eps'})$ time algorithm.}

\paragraph{Why $\neg$ETH does not affect our results.} Very recently, OpenAI~\cite{openai2026eth} and Anthropic~\cite{anthropic2026eth} announced a refutation of the Exponential Time Hypothesis, showing that 3SAT on $n$ variables is solvable in $\exp(n \poly(\log \log n)/\sqrt{\log n})$ time. While this result directly implies that \MaxkSat on $n$ variables and $c n$ clauses is in $2^{o(n)}$ time (refuting the ``sparse'' hypothesis~\cite{DBLP:conf/innovations/AlmanW20}), it evidently says nothing about the full \MaxkSat Hypothesis. This is because \MaxkSat instances may have up to $\Theta(n^k)$ clauses, and \emph{all} known reductions from \MaxkSat on $m$ clauses to $3$SAT introduce $\Omega(m)$ variables. (In other words, no Sparsification Lemma~\cite{DBLP:journals/jcss/ImpagliazzoPZ01} is known for $\MaxkSat$ for any $k$.) Hence, we would apparently need $3$SAT to be in less than $\exp(n^{1/k})$ time in order to solve \MaxkSat faster via reductions. Our results directly solve \MaxkSat faster by solving  $k$-hyperclique faster.

Intriguingly, our algorithm for finding hypercliques faster does share one ingredient with the OpenAI refutation of ETH~\cite{openai2026eth}: namely, they both rely on an interesting new decomposition of the identity matrix. See \Cref{rem:ETH} and the discussion before it.

\subsection{The algebraic approach to hyperclique}

\label{sec:hc}

As mentioned earlier, the case of the $k$-Hyperclique Hypothesis for $k=3$ (in ``$2$-uniform hypergraphs'', i.e., graphs) is false: this is simply the problem of triangle detection in graphs, and it is solved by computing $\mathrm{tr}(A^3)$, the trace of the adjacency matrix cubed. Letting $X$ be an $n \times n$ matrix of $x_{i,j}$'s, we can view the triangle algorithm as evaluating the polynomial
\[\Delta^{(3)}_n = \sum_{i,j,k \in [n]} x_{i,k} x_{k,j} x_{i,j}\] on an appropriate $0/1$ input.

This leaves the tantalizing possibility of tackling the hyperclique problem via a similar algebraic approach, which is usually formulated as the problem of evaluating the \emph{hyperclique polynomial}.

\paragraph{The hyperclique polynomial.} Let $k \geq 2$.
Given a $k$-tuple $\alpha \in [n]^k$ and an index $j \in [k]$, let $\alpha_{-j}$ denote the tuple in $[n]^{k-1}$ with the $j$-th coordinate of $\alpha$ removed. The \emph{hyperclique polynomial} is defined by
\[\Delta^{(k)}_n(X) \coloneqq \sum_{\alpha \in [n]^k} \prod_{j=1}^k x^{(j)}_{\alpha_{-j}}.\]
This is a multilinear polynomial of degree $k$ (also called a \emph{$k$-mode tensor}), with $n^{k-1}$ variables in each variable group
\[X^{(j)} = \{x^{(j)}_\beta , \beta \in [n]^{k-1}\}.\]
Given a $(k-1)$-uniform, $k$-partite hypergraph $H = \bigsqcup_{i=1}^k H_i$ with vertex sets $H_i = [n]$, let $x^{(j)}_{\alpha_{-j}} = 1$ if $H$ contains a $(k-1)$-hyperedge $\alpha_{-j}$ on vertex parts other than $j$, and 0 otherwise. Evaluating $\Delta^{(k)}_n$ at $X$ then counts the number of $k$-hypercliques in $H$.

For example, when $k = 2$ we have $\Delta^{(2)}_n = \sum_{i,j} x_j^{(1)} x_i^{(2)}$. When $k=3$, after renaming the variables and indices, the polynomial $\Delta^{(3)}_n = \sum_{i,j,k} x^{(1)}_{jk} x^{(2)}_{ik} x^{(3)}_{ij}$ becomes the \emph{matrix multiplication tensor}:
\[ \operatorname{tr}(XYZ) = \sum_{i,j,k} x_{ij} y_{jk} z_{ki}. \]

It is clear that $\Delta^{(k)}_n$ has circuits of size $O(n^k)$, as there are $n^k$ monomials in the standard basis. In the case of $k=2$ and $k=3$, polynomial improvements over this baseline are possible: $\Delta^{(2)}_n$ straightforwardly has a circuit of size $O(n)$, and for $k=3$ we have had circuits of size $n^{3 - \varepsilon}$ since work of Strassen \cite{Str69Gaussian}. \Cref{thm:hyperclique} gives polynomial improvements over the trivial bound for all $k$.

\subsection{Arithmetic circuits and tensor ranks}\label{subsec:flattening}

Before proceeding to our improvement for the hyperclique polynomials, we discuss why people had believed the hyperclique conjecture even though there are nontrivial algorithms for matrix multiplication. We begin by reviewing the high-level approach of fast matrix multiplication algorithms.

\paragraph{Tensor rank and its ``barrier''.}

A fundamental measure of a tensor is the \emph{tensor rank} $\Rk(f)$, which describes the minimal number of terms required to write $f$ as a sum of products of linear factors. Formally, $\Rk(f)$ is the smallest $r$ such that the following decomposition exists:
\begin{equation}
    f(X) = \sum_{i=1}^r \ell_i^{(1)}(X^{(1)}) \cdots \ell_i^{(k)}(X^{(k)}), \label{eqn:tensor-rank}
\end{equation}
where $\ell_i^{(j)}$ is a linear function in $X^{(j)}$.

The importance of tensor rank is twofold.
\begin{itemize}
    \item On one hand, in the world of 3-mode tensors, Strassen~\cite{Str73Vermeidung} proved that tensor rank characterizes the \emph{multiplicative} complexity, which is the number of \emph{nonscalar} multiplications required to evaluate the tensor. It thus provides a lower bound for the circuit complexity.
    \item On the other hand, tensor rank upper bounds can be generally turned into algorithms. In fact, the seminal algorithm for matrix multiplication due to Strassen~\cite{Str69Gaussian} can be seen as finding a $7$-term rank decomposition for $2\times 2$ matrix multiplication. This implies an $O(n^{\log_2 7})$-size arithmetic circuit for general $n\times n$ matrix multiplication.
\end{itemize}

One might hope to generalize Strassen's approach to higher-order tensors. The trivial rank decomposition of the hyperclique tensor $\Delta^{(k)}_n$ has $n^k$ terms. Improving this upper bound implies nontrivial algorithms --- in particular, a rank $r$ decomposition for the order-$q$ case where $r < q^k$ would give us an arithmetic circuit of size $O(n^{\log_q r})$, beating the brute force algorithm. Alas, this approach appeared to be impossible: Lincoln, Vassilevska Williams and Williams~\cite[Theorem~8.1]{lincoln2018tight} pointed out that when $k\geq 4$, the trivial $n^k$ tensor rank upper bound for $\Delta^{(k)}_n$ is in fact optimal. They took this as evidence that arithmetic circuits cannot beat brute force for hyperclique, and it was one of their main reasons for proposing the hyperclique conjecture.

\paragraph{Partitions and ranks.}
However, a lower bound on tensor rank doesn't rule out the possibility of having small arithmetic circuits for higher modes. A folklore example is known as the \emph{product of inner products} tensor:
\begin{equation}
    \PIP_n(X, Y, Z, W) = \Biggl(\sum_{i=1}^n x_i z_i\Biggr) \Biggl( \sum_{j=1}^n y_j w_j \Biggr). \label{eqn:pip-decompose}
\end{equation}
It is clear from the definition that $\PIP_n$ has a circuit of size $O(n)$. On the other hand, this tensor has tensor rank $n^2$. The rank lower bound proof is actually quite simple, using an approach called \emph{flattening} which we demonstrate here to give insight into the limitation of tensor rank.

We can expand the expression of the PIP tensor
\begin{equation}
    \PIP_n(X, Y, Z, W) = \sum_{i,j=1}^n x_i y_j \cdot z_i w_j, \label{eqn:pip}
\end{equation}
which is naturally an $n^2$-term rank decomposition. One way to see that $n^2$ is also a lower bound is to \emph{flatten} the tensor: think of this tensor as an $n^2\times n^2$ matrix, by letting the indices of $X, Y$ be the rows, and $Z, W$ be the columns. Every single term appearing in (\ref{eqn:tensor-rank}) would be a rank-$1$ matrix after flattening; this means if a tensor has rank $r$, then the corresponding matrix also has rank at most $r$. However, after flattening, this particular tensor becomes an identity matrix, which has rank $n^2$. This tells us that $n^2$ is also a lower bound on the tensor rank.

Besides the concrete example above, let's again consider another argument which is a structural view of arithmetic circuits. For a tensor, we can actually assume\footnote{This is called the \emph{(set-)multilinearization}~\cite[\S3]{Raz13} of arithmetic circuits. More precisely, it means one can always turn a circuit into that form with only a constant factor blow-up.} the arithmetic circuit evaluating the tensor respects its multilinear structure, meaning every gate computes a multilinear polynomial with respect to a subset of the modes. For product gates, since each gate is computing a tensor, the inputs of a product gate must partition its related modes into several different parts.

Tensor rank can be thought of as corresponding to a restricted class of circuits of the form ``$\Sigma \Pi \Sigma$'' studied by algebraic complexity theorists: There is only one layer of product gates, and since its output should be a $k$-mode tensor, it necessarily should be written as a product of linear forms. However, a general arithmetic circuit can have more product layers, e.g., $\Sigma \Pi \Sigma \Pi \Sigma$, so the top product gates can be products of some genuine multilinear polynomials which are not linear forms. Indeed, the straightforward linear-size circuit for the PIP polynomial is a $\Pi \Sigma \Pi$ circuit, which has two product layers.

That said, the $k=3$ case, where Strassen~\cite{Str73Vermeidung} proved that tensor rank provides a lower bound on circuit size, works merely because in this regime every arithmetic circuit necessarily has the form $\Sigma\Pi\Sigma$ (or can be converted to this form with only a constant-factor loss). This hides the intricacy of product gates in arithmetic circuits, which becomes inevitable when one goes to higher modes.

\begin{figure}[htbp]
\centering
\begin{tikzpicture}[
    x=1cm, y=0.42cm, >=Stealth,
    gate/.style={circle, draw, fill=white, minimum size=5.5mm, inner sep=0pt},
    small gate/.style={gate, minimum size=4.5mm},
    leaf/.style={inner sep=1pt, font=\small},
    block/.style={draw, rounded corners=2pt, fill=white, inner sep=3pt, minimum height=5mm},
    wire/.style={black!70},
  ]
  \newcommand{\cwbranch}[5]{
    \node[gate] (#5) at (1.3,{-(#1+1.5)}) {$\times$};
    \node[leaf] (#5a) at (2.6,{-(#1)}) {$#2_0$};
    \node[gate] (#5s) at (2.6,{-(#1+3)}) {$+$};
    \node[small gate] (#5m1) at (3.9,{-(#1+1.5)}) {$\times$};
    \node at (3.9,{-(#1+3)+0.15}) {$\vdots$};
    \node[small gate] (#5mq) at (3.9,{-(#1+4.5)}) {$\times$};
    \node[leaf] (#5l1) at (5,{-(#1+1)}) {$#3_1$};
    \node[leaf] (#5l2) at (5,{-(#1+2)}) {$#4_1$};
    \node[leaf] (#5l3) at (5,{-(#1+4)}) {$#3_q$};
    \node[leaf] (#5l4) at (5,{-(#1+5)}) {$#4_q$};
    \draw[wire] (#5) -- (#5a.west) (#5) -- (#5s)
      (#5s) -- (#5m1) (#5s) -- (#5mq)
      (#5m1) -- (#5l1.west) (#5m1) -- (#5l2.west)
      (#5mq) -- (#5l3.west) (#5mq) -- (#5l4.west);
    \begin{pgfonlayer}{background}
      \fill[black!8, rounded corners=4pt] (0.85,{-(#1-0.55)}) rectangle (5.45,{-(#1+5.55)});
    \end{pgfonlayer}
    \draw[->, thick, black!60] (5.6,{-(#1+2.5)}) -- (6.4,{-(#1+2.5)});
    \node[block, fill=black!20, anchor=west] (#5p1) at (6.55,{-(#1+2.5)}) {$\mathbf{\MakeUppercase{#2}}$};
    \node[block, anchor=west] (#5p2) at ([xshift=2pt]#5p1.east) {$\mathbf{\MakeUppercase{#3}}\;\mathbf{\MakeUppercase{#4}}$};
    \node[anchor=west] at (8.3,{-(#1+2.5)}) {$#2_0 \cdot \sum_{i=1}^q #3_i #4_i$};
  }
  \cwbranch{0}{x}{y}{z}{A}
  \cwbranch{6.5}{y}{x}{z}{B}
  \cwbranch{13}{z}{x}{y}{C}
  \node[gate] (root) at (0,-8) {$+$};
  \draw[wire] (root) -- (A) (root) -- (B) (root) -- (C);
  \draw[<-, wire] (root.west) -- ++(-0.5,0) node[left] {$\mathrm{cw}_q$};
  \foreach \y in {-5.75,-12.25} \node at (9.2,\y) {$+$};
  \node[anchor=west] at (6.55,-19.2) {$\Longrightarrow\ \PR(\mathrm{cw}_q) \le 3$};
\end{tikzpicture}
\caption{Turn a $\Sigma\Pi\Sigma\Pi$ circuit into a partition rank decomposition}
\label{fig:cw}
\end{figure}
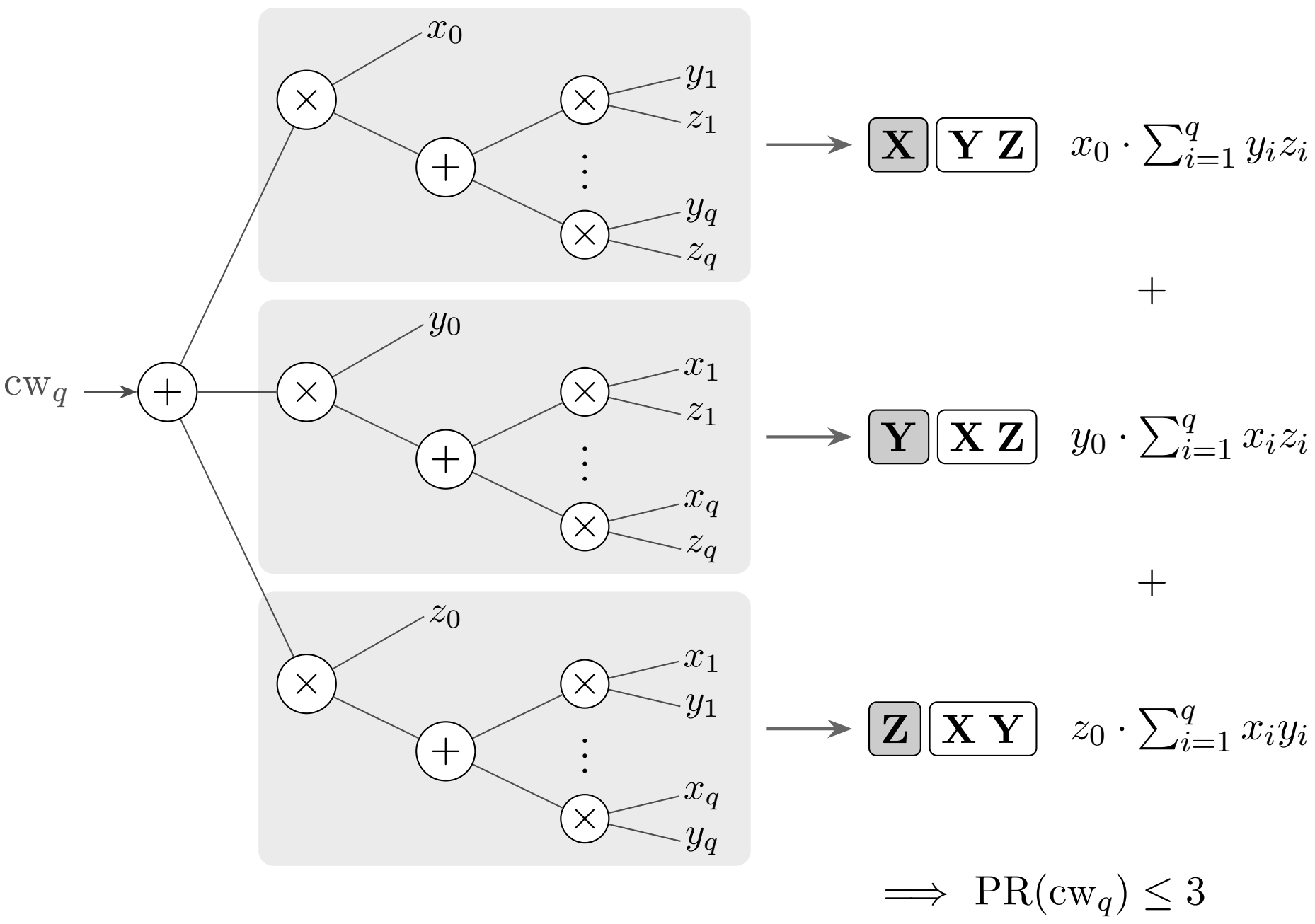

What, then, is the correct notion of rank that can provide useful information for circuit complexity? A natural approach is to make the required terms in a rank decomposition more flexible, so that it's possible to charge each term to a specific product gate. For example, one might consider the \emph{partition rank} defined by Naslund~\cite{Nas20}: $\PR(f)$ is the minimal $r$ needed to write $f$ as a sum of products
\[ f = \sum_{i=1}^r u_i \cdot v_i, \]
where each $u_i$ is a multilinear polynomial in a nonempty and proper subset $S_i$ of the $k$ modes, and $v_i$ is a multilinear polynomial in the remaining modes $[k] \setminus S_i$. For example, \Cref{fig:cw} shows how to turn a circuit evaluating the (small) Coppersmith--Winograd tensor $\mathrm{cw}_q = \sum_{i=1}^q (x_0 y_iz_i + x_iy_0z_i + x_i y_i z_0)$ into a 3-term partition rank decomposition.

The partition rank is actually a poor measure of circuit complexity --- one can always write a tensor as $f = \sum_i x_i^{(1)}\cdot f_i$ where $f_i$ is the $x_i^{(1)}$-term coefficient, so partition rank can at most be linear\footnote{It is often \emph{sublinear}, which has found applications in combinatorics~\cite{CLP17,EG17,TS16SliceRank,NS17,Nas20}.}, which cannot provide a meaningful bound for circuit size. However, it turns out that one can address this by studying some restricted notions of partition ranks, in which we only allow certain kinds of partitions $S_i$ in the decomposition.

These kinds of partition ranks appear to admit interesting upper bounds and further guide genuine circuit construction. For example, the study of Hrube{\v{s}}~\cite{hrubevs2026subquadratic} on sum-of-squares identities can be interpreted as bounding the restricted partition rank of the PIP tensor: One can decompose $\PIP_n$ into a sum of $O(n^{1.62})$ many $(XY|ZW)$ and $(XW|YZ)$ terms\footnote{Note that $(XZ|YW)$ terms shouldn't be allowed, as \eqref{eqn:pip-decompose} will just give a decomposition with rank 1!}. He uses these methods to construct an interesting \emph{noncommutative} circuit, of size only $O(n^{1.96})$, computing $\PIP_n$ when it is defined as in \eqref{eqn:pip} with noncommuting variables.

Our presentation of the circuit for the hyperclique polynomial will follow a similar route: We will first explain how to obtain a certain kind of partition rank of $\Delta^{(k)}_n$, then eventually extend this partition rank upper bound to a circuit size bound.

We also remark that recently, Brand, Kaski and Wang~\cite{brand2026partition} proposed another modification of partition ranks, which aims to generalize Strassen's lower bound to higher modes. They define the $(k-2,1,1)$-\emph{partition rank} as the minimal number of terms needed to write a tensor as a sum of products of a $(k-2)$-mode multilinear polynomial, a linear form and another linear form. This notion $\Rk_{k-2,1,1}(f)$ can provide a lower bound for arithmetic circuits. In principle, to find a small circuit for a tensor, one should first find a small $(k-2,1,1)$-partition rank decomposition. (Although we will not exactly follow this as it's cumbersome to explicitly write down such a decomposition from our circuits.)

\subsection{A partition rank decomposition} \label{sec:pr}

In this subsection, we will focus on the $k=4$ case for simplicity, and prove a certain kind of $O(n^{4-\varepsilon})$-term partition rank upper bound of the hyperclique polynomial. This can be seen as a first step towards the algebraic refutation of the hyperclique conjecture: We are partially constructing an arithmetic circuit by first building its top product layer, then the completion of the rest of the circuit will be sketched in \cref{sec:overview}.

\paragraph{Merely below brute-force.}
We first begin with a ``baby step,'' finding a decomposition with fewer than $n^4$ terms that does not quite give us a polynomial saving.

For convenience, we will relabel the variables in the following way:
\[ \Delta^{(4)}_n(X,Y,Z,W) = \sum_{i,j,k,\ell = 1}^n x_{j k \ell} y_{i k \ell} z_{i j \ell} w_{i j k}. \]
We strategically first pair the variable set into groups $XZ$ and $YW$:
\[ \Delta^{(4)}_n(X,Y,Z,W) = \sum_{i,j,k,\ell = 1}^n x_{j k \ell} z_{i j \ell} \cdot y_{i k \ell} w_{i j k}, \]
then we artificially add and then subtract a term $x_{i k \ell} z_{j i \ell}$; note that this is interchanging $i$ and $j$ for $x_{j k \ell} z_{i j \ell}$.
\begin{align*}
    \Delta^{(4)}_n(X,Y,Z,W) &= \sum_{i,j,k,\ell = 1}^n (x_{j k \ell} z_{i j \ell} - x_{i k \ell} z_{j i \ell} + x_{i k \ell} z_{j i \ell}) \cdot y_{i k \ell} w_{i j k}\\
    &= \sum_{i,j,k,\ell = 1}^n (x_{j k \ell} z_{i j \ell} - x_{i k \ell} z_{j i \ell})  y_{i k \ell} w_{i j k}
    + \sum_{i,j,k,\ell = 1}^n x_{i k \ell} z_{j i \ell} \cdot y_{i k \ell} w_{i j k}\\
    &\eqqcolon T_0 + T_1.
\end{align*}
The advantage comes from analyzing each of $T_0$ and $T_1$ separately. For $T_0$, the saving comes from the fact that $x_{j k \ell} z_{i j \ell} - x_{i k \ell} z_{j i \ell}$ is \emph{antisymmetric} with respect to $i$ and $j$, meaning that swapping them negates the term. Thus we can write it as
\[ T_0 = \sum_{1\leq i < j\leq n} \sum_{k,\ell=1}^n (x_{j k \ell} z_{i j \ell} - x_{i k \ell} z_{j i \ell}) (y_{i k \ell} w_{i j k} - y_{j k \ell} w_{j i k}), \]
which only has $\binom{n}{2} n^2 \approx n^4 / 2$ terms, a factor of $2$ below the straightforward $n^4$. 

The saving for $T_1$ comes from the change in \emph{which indices each variable depends on}. Looking at the indices of the four variables in
\[ T_1 = \sum_{i,j,k,\ell = 1}^n x_{i k \ell} z_{j i \ell} \cdot y_{i k \ell} w_{i j k}, \]
we see that compared to $\Delta^{(4)}_n$, our manipulation didn't change the indices that $z$ depends on (it still depends on $j, i, \ell$, although read in a different order), but it changed the indices that $x$ depends on: it now depends on $i, k, \ell$ rather than on $j, k, \ell$. Consequently, every variable depends on $i$ now, and after rearrangement,
\begin{align*}
    T_1 &= \sum_{i=1}^n \sum_{j,k,\ell = 1}^n (x_{i k \ell} \cdot y_{i k \ell}) z_{j i \ell}  w_{i j k}\\
    &= \sum_{i=1}^n \Delta^{(3)}_n(\{x_{i k \ell} \cdot y_{i k \ell}\}_{k\ell}, \{z_{j i \ell}\}_{j \ell}, \{w_{i j k}\}_{j k}),
\end{align*}
which becomes a sum of $n$ invocations of the $\Delta^{(3)}_n$ tensor! Since we know $\Delta^{(3)}_n$ has tensor rank $n^{\omega + o(1)}$ with $\omega < 3$, after substituting $x\cdot y, z, w$ into it, every term in the tensor rank decomposition becomes an $(XY | Z | W)$-partition rank term.

Putting these decompositions together, we now have a way to rewrite $\Delta^{(4)}_n$ as $T_0 + T_1$, where
\begin{itemize}
    \item $T_0$ has about $n^4 / 2$ many $XZ|YW$-partition rank terms, and
    \item $T_1$ has $n^{\omega + 1 + o(1)}$ many $XY | Z | W$-partition rank terms.
\end{itemize}
Because $\omega < 3$, the first term dominates, and we have a certain kind of partition rank that has roughly $n^4 / 2$ terms, bypassing the tensor rank bound.

We can sanity check this decomposition by comparing it with the rank lower bound from flattening that we discussed in \cref{subsec:flattening} above. That argument gives an $n^4$ lower bound for any partition into two parts on each side, such as $(XZ|YW)$. Our decomposition has terms partitioned into both $(XZ|YW)$ and $(XY | Z | W)$, and no flattening with two parts on each side would convert each such term into a rank 1 matrix. 
More generally, for $k\geq 4$, the $n^k$-term tensor rank lower bound is proved via any non-singleton flattening (which means the flattening must have at least two modes on both sides).

\paragraph{Tensor up?}

A common strategy to boost a nontrivial bound to a larger polynomial saving is through taking \emph{tensor powers}. Let $f, g$ be two $k$-mode tensors
\[ f = \sum_{i_1,\dots,i_k} f_{i_1,\dots,i_k} x_{i_1}^{(1)} \cdots x_{i_k}^{(k)}, \quad
   g = \sum_{j_1,\dots,j_k} g_{j_1,\dots,j_k} x_{j_1}^{(1)} \cdots x_{j_k}^{(k)}. \]
Their \emph{tensor product} (also called \emph{Kronecker product}) is defined as
\[ f\otimes g = \sum_{\substack{i_1,\dots,i_k \\ j_1,\dots,j_k}} f_{i_1,\dots,i_k} g_{j_1,\dots,j_k} \cdot x_{i_1j_1}^{(1)} \cdots x_{i_kj_k}^{(k)}. \]
A notable property of tensor rank is its \emph{submultiplicativity} $\Rk(f \otimes g) \leq \Rk(f) \cdot \Rk(g)$. This is because, if $f, g$ can be written as a sum of products
\[ f = \sum_i \ell_i^{(1)} \cdots \ell_i^{(k)}, \quad g = \sum_j \ell_j^{(1)} \cdots \ell_j^{(k)}, \]
then their tensor product can as well by expanding
\begin{align*}
    f\otimes g &= \sum_{i,j} (\ell_i^{(1)} \cdots \ell_i^{(k)}) \otimes (\ell_j^{(1)} \cdots \ell_j^{(k)})\\
    &= \sum_{i,j} (\ell_i^{(1)} \otimes \ell_j^{(1)}) \cdots (\ell_i^{(k)} \otimes \ell_j^{(k)}),
\end{align*}
because the tensor product of rank-1 terms is again a rank-1 term.

Hyperclique tensors have the self-similarity property $\Delta^{(k)}_{nm} = \Delta^{(k)}_n \otimes \Delta^{(k)}_m$. This allows one to bound the tensor rank for the whole family via a finite construction, especially for the $k = 3$ case: If one finds a tensor rank bound $\Rk(\Delta_q^{(3)}) \leq r$, then for $n = q^d$ we have
\[ \Rk(\Delta_{q^d}^{(3)}) = \Rk\bigl((\Delta_{q}^{(3)})^{\otimes d}\bigr) \leq r^d. \]
With some more care, one can use this to show the rank and the circuit complexity of $\Delta_n^{(3)}$ are both $O(n^{\log_q r})$, which means $\omega \leq \log_q r$.

Can the tensor product argument also let us boost our ``partition rank'' estimate from $< n^4$ to $n^{4-\varepsilon}$? Unfortunately, the most direct approach doesn't work, since our partition rank decomposition contains different partitions. We just showed $\Delta_n^{(4)} = T_0 + T_1$, where $T_0$ is a sum of $XZ|YW$ partitions and $T_1$ is a sum of $XY | Z | W$ partitions. Suppose we start with some fixed-$q$ case $\Delta_q^{(4)} = T_0 + T_1$ of this decomposition, and want to use it to bound the tensor power $\Delta_{q^d}^{(4)} = (\Delta_q^{(4)})^{\otimes d}$. For a subset $A\subseteq [d]$, write $T_A = T_0^{\otimes [d] \setminus A} T_1^{\otimes A}$ to denote the tensor product of $d$ factors, where the $a$-th tensor factor is $T_1$ if $a\in A$ and $T_0$ otherwise. Expanding the tensor power gives
\[ \Delta_{q^d}^{(4)} = (T_0 + T_1)^{\otimes d} = \sum_{A\subseteq [d]} T_A. \]
The cases $A = \emptyset$ and $A = [d]$ are well-behaved, because $T_0^{\otimes d}$ ($T_1^{\otimes d}$, resp.) remains an $XZ|YW$ ($XY|Z|W$, resp.) partition rank decomposition. The more challenging part is the mixed case, since the tensor product $T_0 \otimes T_1$ of our partitions is neither an $XZ|YW$ nor an $XY|Z|W$ partition.

\paragraph{Dichotomy.}
Nonetheless, this problem can be resolved by a \emph{dichotomy} argument. Intuitively, we will argue that if $|A|$ is small, then the mixed tensor product $T_A$ has a lot of $T_0$ factors and thus behaves like $T_0^{\otimes d}$; we refer to this as the \emph{low-degree} case, and notationally will label tensors corresponding to this case with $L$, since they will correspond to the case in \Cref{eq:HL} below with low degree in the $\tau$ variables. Similarly, if $|A|$ is large, then $T_A$ intuitively behaves like $T_1^{\otimes d}$, and we call this the \emph{high-degree} case labeled by $H$.

We pick some threshold $t$ to be determined later, and consider splitting the summation into
\[ (T_0 + T_1)^{\otimes d} = \sum_{|A| \leq t} T_A + \sum_{|A| > t} T_A \eqqcolon \Delta^{(4)}_{q^d, L} + \Delta^{(4)}_{q^d, H}. \]
We can actually argue that the low-degree part $\Delta^{(4)}_{q^d, L}$ still has small $XZ|YW$ partition rank and the high-degree part $\Delta^{(4)}_{q^d, H}$ has small $XY|Z|W$ partition rank.

\emph{Low-degree.} The $XZ|YW$ partition rank of a tensor is just the matrix rank of the flattening in which pairs of $x$ and $z$ variables become the rows, and pairs of $y$ and $w$ variables become the columns. The row space of the original tensor $\Delta_q^{(4)}$ is
\[ V = \Span \{ x_{jk\ell} z_{ij\ell} \}_{i,j,k,\ell \in [q]}, \]
while the row space of $T_0$ lives in the subspace antisymmetrizing $i$ and $j$:
\[ V_0 = \Span \{ x_{jk\ell} z_{ij\ell} - x_{ik\ell} z_{ji\ell} \}_{i,j,k,\ell \in [q]} \subset V. \]
Think of $V$ as decomposed into $V_0$ and its complement, $V = V_0 \oplus V_1$. Then any low-degree $T_A$ with $|A| \leq t$ lives in the space where at most $t$ tensor factors are $V_1$, which gives the rank estimate for $\Delta^{(4)}_{q^d, L}$ of
\begin{equation}
    \Rk_{XZ|YW}(\Delta^{(4)}_{q^d, L}) \leq \sum_{b\leq t} \binom d b (\dim V_0)^{d-b} (\dim V_1)^b = \sum_{b\leq t} \binom d b \Bigl[\frac{q^3(q-1)}{2}\Bigr]^{d-b} \Bigl[\frac{q^3(q+1)}{2}\Bigr]^b. \label{eqn:pr-low}
\end{equation}

\emph{High-degree.} For the $XY|Z|W$ partition rank, we directly use the submultiplicativity of rank to bound each $T_A$ for any high-degree $|A| = b > t$ by
\[ \Rk_{XY|Z|W}(T_A) \leq \Rk_{XY|Z|W}(T_0)^{d-b} \Rk_{XY|Z|W}(T_1)^b. \]
Write $r = \Rk_{XY|Z|W}(T_1)$. Since $T_0 = \Delta_{q}^{(4)} - T_1$, we can also bound $\Rk_{XY|Z|W}(T_0)$ by $q^4 + r$. This gives the rank estimate for $\Delta^{(4)}_{q^d, H}$
\begin{equation}
    \Rk_{XY|Z|W}(\Delta^{(4)}_{q^d, H}) \leq \sum_{b > t} \binom{d}{b} (q^4 + r)^{d-b} r^b. \label{eqn:pr-high}
\end{equation}

\emph{Threshold.} We want to set the threshold $t$ appropriately such that both parts of the partition rank estimate are $(q^d)^{4-\varepsilon}$.

For the low-degree part, note that the whole summation \eqref{eqn:pr-low} just sums to $q^{4d}$ when there's no threshold $t$. Most of the mass concentrates at the summands where $b / d \approx (\dim V_1) / (\dim V) = (q+1)/2q$, so to get exponential saving, one only needs to pick $t$ below this; we may just set $t = d/2$.

For the high-degree part, note that we have $r = q^{\omega + 1 + o(1)}$ where $\omega < 3$, so $r / q^4 \to 0$ when $q \to \infty$. When $q$ is large enough, most of the contribution of \eqref{eqn:pr-high} just comes from the $b \approx d/2$ case, which is roughly $[2\sqrt{(q^4+r)\cdot r}]^d$. Again, when $q$ is sufficiently large, this is smaller than $q^{4d}$. Thus $\Delta^{(4)}_{q^d, L}$ and $\Delta^{(4)}_{q^d, H}$ have partition rank $q^{(4-\varepsilon)d}$.

We remark that dichotomy arguments like this have already appeared in previous studies of tensors and matrices, including some partition rank (including slice rank) upper bounds in combinatorics~\cite{CLP17,EG17,TS16SliceRank,NS17,Nas20}, as well as matrix rigidity upper bounds~\cite{AW17,Alm21}.

In the next subsection, we take this partition rank construction and complete it into a genuine circuit. Before proceeding, we remark on a few technical differences which will arise as we do this. For the general analysis of the high-degree part $\Delta^{(k)}_{q^d, H}$, we will be able to almost straightforwardly replace the tensor rank bound \eqref{eqn:pr-high} by a circuit bound, since our approach mainly reduces it to many $(k-1)$-hyperclique polynomials. For the low-degree part $\Delta^{(k)}_{q^d, L}$, we will have more work to do, particularly designing circuits computing the two branches of the partition rank decomposition. This will require us to lower the threshold to $t = d / B$ for some constant $B > 2$ depending on $k$.

\subsection{Circuit construction} \label{sec:overview}
In this subsection, we will sketch how to extend the previous partition rank upper bound to a circuit size upper bound; in \cref{sec:mainproof} we will formally prove the result sketched here. The proof proceeds by induction on $k$: we assume that $\Delta^{(k-1)}_n$ has circuits of size $O(n^{k-1-\varepsilon})$ for some $\varepsilon > 0$, and show that this implies that $\Delta^{(k)}_n$ has a circuit of size $n^{k - \varepsilon'}$ for some $\varepsilon' > 0$. We start from the base case that $\Delta^{(3)}_n$ has circuits of size $n^{\omega + o(1)}$.

From now on, let $k$ and $\varepsilon$ be fixed. We will write $n = q^d$ with $q = 2^B$, and $B$ will be a large constant depending on $k$ and $\varepsilon$. If $n$ cannot be expressed in this way, pad the instance with extra vertices, which only changes the circuit sizes by a constant.

\paragraph{Setup.}
We start by introducing some notation. Label the $k$ coordinates as $(i,j,z_1, \ldots, z_{k-2}) \in [n]^k$. Introduce variables $U(i,z), V(j,z)$, and $X_s(i,j,z_{-s})$ for $i,j \in [n], z \in [n]^{k-2}, s \in [k-2]$. Then we can equivalently write the $k$-hyperclique tensor as the polynomial
\[\Delta^{(k)}_n = \sum_{i,j \in [n], z \in [n]^{k-2}} U(i, z) V(j,z) \prod_{s=1}^{k-2}X_s(i,j,z_{-s}). \]

We partition the $k$ modes by making $U$ and $X_{k-2}$ into one group and the rest of the modes into another group. Define matrices $Q^z$ and $P^z$ via
\begin{align*}
(Q^z)_{ij} &= U(i,z) X_{k-2}(i,j,z_{-(k-2)}),\\
(P^z)_{ij} &= V(j,z) \prod_{s=1}^{k-3} X_{s}(i,j,z_{-s}).
\end{align*}
These are $n \times n$ matrices with monomial entries. We then have 
\[\Delta^{(k)}_n = \sum_z \langle Q^z, P^z \rangle\]
where $\langle A,B \rangle = \sum_{i,j} A_{ij}B_{ij}$ is the Frobenius inner product. So far we have done nothing other than rename and group variables.

For a linear map $T$ on the space of $n \times n$ matrices, define the \emph{twisted} hyperclique tensor
\[\Delta^{(k)}_{n,T} \coloneqq \sum_{z} \langle P^{z}, T \cdot Q^{z} \rangle.\]
We next describe how to rephrase the decomposition $\Delta^{(k)}_n = \Delta^{(k)}_{n,I} = \Delta^{(k)}_{n, L} + \Delta^{(k)}_{n, H}$ discussed in \Cref{sec:pr} in terms of linear operators $L, H$ such that $I = L+H$.

\paragraph{Partial transposes.}
We will view $q^d \times q^d$ matrices as having rows and columns indexed by vectors in $[q]^d$. For $p \in [d]$ we define the transpose operator $\tau_p$, which acts on the space of $q^d \times q^d$ matrices via

\[\tau_p \cdot Q_{(a_1, \ldots, a_p, \ldots,  a_d), (b_1, \ldots, b_p, \ldots, b_d)}  = Q_{(a_1, \ldots, b_p, \ldots,  a_d), (b_1, \ldots, a_p, \ldots, b_d)}.\]

Note that transpose operators commute with each other, so for $A \subseteq [d]$ there is no ambiguity in defining $\tau_A \coloneqq \prod_{a \in A} \tau_a$. We call these $\tau_A$ \emph{partial transposes}. These operators appear frequently in quantum information theory \cite[Section 6.2.3]{watrous2018theory}.

Similar to the analysis of high-degree part in \Cref{sec:pr},
let's see how partial transposes allow us to reduce $\Delta^{(k)}_{n,\tau_A}$ to $\Delta^{(k-1)}_n$. Consider first the extreme case of $\Delta^{(k)}_{n,\tau_{[d]}}$, where we fully transpose the $Q^z$ matrices. We have that
\[\Delta^{(k)}_{n,\tau_{[d]}} =  \sum_z \langle P^{z}, \tau_{[d]} \cdot Q^{z} \rangle = \sum_z \langle P^{z}, (Q^{z})^\sfT \rangle =\sum_{i,j,z} V(j,z) \prod_{s=1}^{k-3} X_{s}(i,j,z_{-s}) \cdot U(j,z) X_{k-2}(j,i,z_{-(k-2)}).\]
Now all subscripts involve $j$. If we first sum over $j$, we have
\begin{align*}
\Delta^{(k)}_{n,\tau_{[d]}} &= \sum_j \sum_{i,z} V(j,z) \prod_{s=1}^{k-3} X_{s}(i,j,z_{-s}) \cdot U(j,z) X_{k-2}(j,i,z_{-(k-2)}) \\
& = \sum_j \sum_{i,z} W(j,z)  \cdot X_{k-2}(j,i,z_{-(k-2)}) \cdot \prod_{s=1}^{k-3} X_{s}(i,j,z_{-s})
\end{align*}
where $W(j,z) \coloneqq U(j,z)V(j,z)$. 
The key observation is that for fixed $j$, each variable in the inner sum omits one subscript among $i,z_1, \ldots, z_{k-2}$. Hence the inner sum equals $\Delta^{(k-1)}_n(W,X^{(1)}, \ldots, X^{(k-2)}).$ So for each $j \in [n]$, the inner sum can be computed using $n^{k-2}$ multiplications to compute the $W$ variables, and then using a circuit for $\Delta^{(k-1)}_n$. Adding up the results for all $j \in [n]$, this gives a circuit of size
\[n\cdot (n^{k-1-\varepsilon} + n^{k-2}) = O(n^{k-\varepsilon}).\]

Hence, for a full transpose, we inherit the savings $n^{-\varepsilon}$ from the $k-1$ case. For a partial transpose $\tau_A$, we can first enumerate over all digits outside $A$, each subproblem becomes a full transposed case for $n = q^{|A|}$, which gives a saving $q^{-\varepsilon |A|}$ in general.

\paragraph{Dichotomy.} We then rephrase the decomposition in \Cref{sec:pr} in terms of linear operators by first writing
\[ I = \prod_{a \in [d]} (\tau_a + (I- \tau_a)) = \sum_{A \subseteq [d]} \left(\prod_{a \in A} \tau_a\right) \left(\prod_{a \notin A} (I-\tau_a)\right). \]

For a threshold $t$ on the number of transpose factors, which we will choose later, break this sum into low and high degree parts:
\begin{align} \label{eq:HL}
L &= \sum_{|A| \le t} \left(\prod_{a \in A} \tau_a\right) \left(\prod_{a \notin A} (I-\tau_a)\right)\\
H &= \sum_{|A| > t} \left(\prod_{a \in A} \tau_a\right) \left(\prod_{a \notin A} (I-\tau_a)\right) = I - L
\end{align}
If we expand $H$ as a polynomial in the $\tau_a$'s, every term has degree greater than $t$. Also, every term in $L$ does ``antisymmetrization'' $I - \tau_a$ to at least $d-t$ coordinates. 

\begin{remark}\label{rem:ETH}
A similar decomposition of the identity matrix (writing $I$ as the sum of a ``low rank'' matrix and a very sparse and structured matrix) also appears in the recently-announced OpenAI algorithm for refuting ETH~\cite{openai2026eth}. 
\end{remark}

By an extension of the previous argument about the full transpose $\tau_{[d]}$, $\Delta^{(k)}_{n,H}$ has small circuits when $t$ is a sufficiently large linear function of $d$ (we will ultimately pick $t = \lfloor d/B \rfloor$). This is our \Cref{cor:finalhigh}.  The remaining challenge is to give a small circuit for $\Delta^{(k)}_{n,L}$ in this regime.

\paragraph{Low-degree.}
To compute $\Delta^{(k)}_{n,L}$, we will work in the basis of $q^d \times q^d$ matrices given by the eigenvectors of the transpose operators. Let $\calB$ be the basis for the space of $q \times q$ matrices given by
\[\mathcal{B} = \{e_{ii}\}_{i \in [q]} \cup \{e_{ij} + e_{ji}, e_{ij} - e_{ji}\}_{1 \le i < j \le q}.\] Then $\mathcal{B}^{\otimes d}$ is a basis for the space of $q^d \times q^d$ matrices; this is an orthogonal basis in which each $\tau_A$ acts diagonally with eigenvalues $\pm 1$ (\Cref{lem:basis}). For $\sigma = (\sigma_1, \ldots, \sigma_d)$ with $\sigma_i \in \mathcal{B}$, we let $E_\sigma = \bigotimes_{i \in [d]} \sigma_i$. Note that $E_\sigma$ has at most one nonzero entry in each row and column, and its nonzero entries are $\pm 1$. Let $\pi_i(\sigma)$ denote the column index of a nonzero entry in row $i$ in $E_\sigma$. Since the support of $E_\sigma$ is symmetric, $\pi_j(\sigma)$ also denotes the row index of the nonzero entry in column $j$. Let $u(\sigma)$ be the number of non-diagonal factors in $\sigma$, and let $a(\sigma)$ be the number of skew-symmetric factors in $\sigma$ (so $a(\sigma) \le u(\sigma))$.

Because $\calB^{\otimes d}$ is an orthogonal basis that diagonalizes $L$, we have
\[\Delta^{(k)}_{n,L} = \sum_z \langle P^z, LQ^z \rangle = \sum_{z, \sigma : a(\sigma) \ge d-t} c_\sigma \langle P^z, E_\sigma\rangle \langle Q^z, E_\sigma \rangle\]
for some (explicit) constants $c_\sigma$; it is not hard to show that any terms with $a(\sigma) < d-t$ vanish. Our goal now is to compute the scalars
\[\mathcal{P}_{\sigma,z}\coloneqq \langle P^z, E_\sigma \rangle, \hspace{1cm} \mathcal{Q}_{\sigma,z} \coloneqq \langle Q^z, E_\sigma \rangle\]
for all $\sigma, z$. We first consider the problem of computing the $\mathcal{Q}$'s. By definition,
\begin{equation}\label{eq:qmotiv}
\calQ_{\sigma,z} = \sum_{i : \row_i(E_\sigma) \neq 0} (Q^z)_{i,\pi_i(\sigma)} (E_\sigma)_{i,\pi_i(\sigma)} = \sum_{i:\row_i(E_\sigma) \neq 0} (E_\sigma)_{i,\pi_i(\sigma)} X_{k-2}(i, \pi_i(\sigma), z_{-(k-2)}) U(i,z).
\end{equation}

\paragraph{Attempt \#1.}
Since $E_\sigma$ only has $2^{u(\sigma)} \le 2^d \le n$ nonzero entries, we might first hope that sparsity already gives a small circuit for the $\calQ_{\sigma,z}$. Doing so we would need 
\[\sum_{z, \sigma : a(\sigma) \ge d-t} \|E_{\sigma}\|_0 = \sum_{z, \sigma : a(\sigma) \ge d-t} 2^{u(\sigma)} = n^{k-2} \sum_{\sigma : a(\sigma) \ge d-t} 2^{u(\sigma)}\]
operations. Unfortunately, just the contribution to this sum for the $\sigma$'s with $a(\sigma) = d-t$ is at least
\[\binom{d}{d-t}  \Bigl[ 2\binom{q}{2} \Bigr]^{d-t} \Bigl[2\binom{q}{2} + q\Bigr]^t \approx \binom{d}{t} n^2\]
which is polynomially larger than $n^2$, since $t$ must be taken linear in $d$ to get savings from the high degree argument (\Cref{lem:highswap}).

However, note that there is redundant computation: many $\calQ_{\sigma,z}$'s will depend on the same entries of $Q^z$. The key insight is that instead of paying cost 1 to compute each term in \eqref{eq:qmotiv} individually for each choice of $\sigma$ and $z$, we should use matrix multiplication to compute many terms simultaneously.

\paragraph{Attempt \#2.}
Let $z = (z', w)$. Our goal of computing the $\calQ_{\sigma,z}$ is equivalent to computing for all $z', w,$ and $\sigma$ with $a(\sigma) \ge d-t$,
\begin{equation}\label{eq:qfeat}
\calQ_{\sigma, (z', w)} = \sum_{i: \row_i(E_\sigma) \neq 0} (E_\sigma)_{i,\pi_i(\sigma)} X_{k-2}(i, \pi_i(\sigma), z') U(i,z',w).
\end{equation}

After fixing $z'$, computing $\calQ_{\sigma, (z', w)}$ for all $\sigma, w$ can be regarded as a matrix multiplication
\[\bigl((E_\sigma)_{i,\pi_i(\sigma)}  X_{k-2}(i, \pi_i(\sigma), z')\bigr)_{\sigma, i} \cdot \bigl(U(i,z',w)\bigr)_{i,w}. \]
How fast should we compute this matrix multiplication in order to get an improvement? The catch is that this matrix multiplication problem is a \emph{sparse} instance --- the first matrix $((E_\sigma)_{i,\pi_i(\sigma)} X_{k-2}(i, \pi_i(\sigma), z'))_{\sigma, i}$ is sparse since every row $\sigma$ only has $2^{u(\sigma)}$ nonzero entries, while the second matrix $(U(i,z',w))_{i,w}$ is dense.

Similar to the estimate of the number of terms in the previous attempt, when we take $t = d/B$, a straightforward calculation shows that the first matrix has $m=n^{2+o(1/B)}$ nonzero entries. Thus this matrix multiplication has $m\cdot n = n^{3 + o(1/B)}$ contributing terms in total. As we need to repeat for all $z' \in [n]^{k-3}$, any polynomial saving in the matrix multiplication will give an improvement. 

To compute the matrix multiplication faster, we can partition the first matrix into many dense matrices. Suppose we can find a set $S$ of $\sigma$'s and a set of column indices $J$ such that each $\sigma \in S$ has a nonzero entry in each $i \in J$. Then we can compute all corresponding terms in \Cref{eq:qfeat} by performing a $|S| \times |J| \times n$ matrix multiplication. If we can \emph{partition} the first matrix this way, such that $|S|, |J| = n^{\Omega_k(1)}$, we will obtain a polynomial saving per matrix multiplication instance, thus a saving for the whole problem.

\begin{figure}[htbp]
\centering
\begin{tikzpicture}[
    x=1cm, y=0.8cm,
    vtx/.style={circle, draw, inner sep=0pt, minimum size=2.2mm},
    sig/.style={vtx, fill=black!25},
    col/.style={vtx, fill=white},
    wvtx/.style={vtx, fill=black!60},
    bA/.style={black, line width=0.9pt},
    bB/.style={black!55, line width=0.9pt, dashed},
    bC/.style={black!80, line width=1pt, densely dotted},
    U/.style={black!20, line width=0.4pt},
  ]

  \foreach \i in {1,...,6} {
    \node[sig] (s\i) at (0,{-(\i-1)}) {};
    \node[col] (j\i) at (3.6,{-(\i-1)}) {};
  }
  \foreach \i in {1,...,4} \node[wvtx] (w\i) at (7,{-0.25-1.5*(\i-1)}) {};
  \begin{scope}[on background layer]
    \foreach \i in {1,...,6} \foreach \k in {1,...,4} \draw[U] (j\i) -- (w\k);
  \end{scope}
  \foreach \a in {1,2,3} \foreach \b in {1,2,3} \draw[bA] (s\a) -- (j\b);
  \foreach \a in {3,4} \foreach \b in {4,5} \draw[bB] (s\a) -- (j\b);
  \foreach \a in {5,6} \foreach \b in {4,5,6} \draw[bC] (s\a) -- (j\b);
  \draw[line width=2.2pt] (s6) -- (j6) -- (w4);
  \draw[line width=1.2pt, dashed, ->, >=Stealth] (s6) to[out=-35, in=-150]
    node[midway, below, font=\small] {contributes to $\calQ_{\sigma,(z',w)}$} (w4);
  \node[above=4pt, font=\small, align=center] at (s1.north) {$\sigma$\\[-2pt]{\footnotesize $a(\sigma)\ge d-t$}};
  \node[above=4pt, font=\small, align=center] at (j1.north) {$i$\\[-2pt]{\footnotesize $[n]$}};
  \node[above=4pt, font=\small, align=center] at (7,0) {$w$\\[-2pt]{\footnotesize $[n]$}};
  \node[font=\footnotesize, align=center] at (1.8,1.9) {$G$ (sparse)\\$(E_\sigma)_{i,\pi_i(\sigma)} \,X_{k-2}(i,\pi_i(\sigma),z')$};
  \node[font=\footnotesize, align=center] at (5.3,1.9) {complete\\$U(i,z',w)$};
  \begin{scope}[shift={(8.2,-0.2)}, font=\small]
    \draw[bA] (0,0) -- (0.8,0) node[right, black] {$S_1\times J_1$};
    \draw[bB] (0,-0.6) -- (0.8,-0.6) node[right, black] {$S_2\times J_2$};
    \draw[bC] (0,-1.2) -- (0.8,-1.2) node[right, black] {$S_3\times J_3$};
    \node[anchor=north west, text width=5.6cm, font=\footnotesize, align=left] at (-0.1,-1.7) {
      Each triangle $(\sigma,i,w)$ is one term of $\calQ_{\sigma,(z',w)}$ in \eqref{eq:qfeat}.
      For each biclique $S_\beta\times J_\beta$, the triangles through it form a
      $|S_\beta|\times|J_\beta|\times n$ matrix product.\\[4pt]
      Total number of triangles:
      \[\textstyle\sum_\beta |S_\beta|\,|J_\beta|\cdot n \;=\; m\cdot n \;\approx\; n^{3}.\]
    };
  \end{scope}
\end{tikzpicture}
\caption{Computing the $\calQ_{\sigma,(z',w)}$ for a fixed $z'$}
\label{fig:Q}
\end{figure}
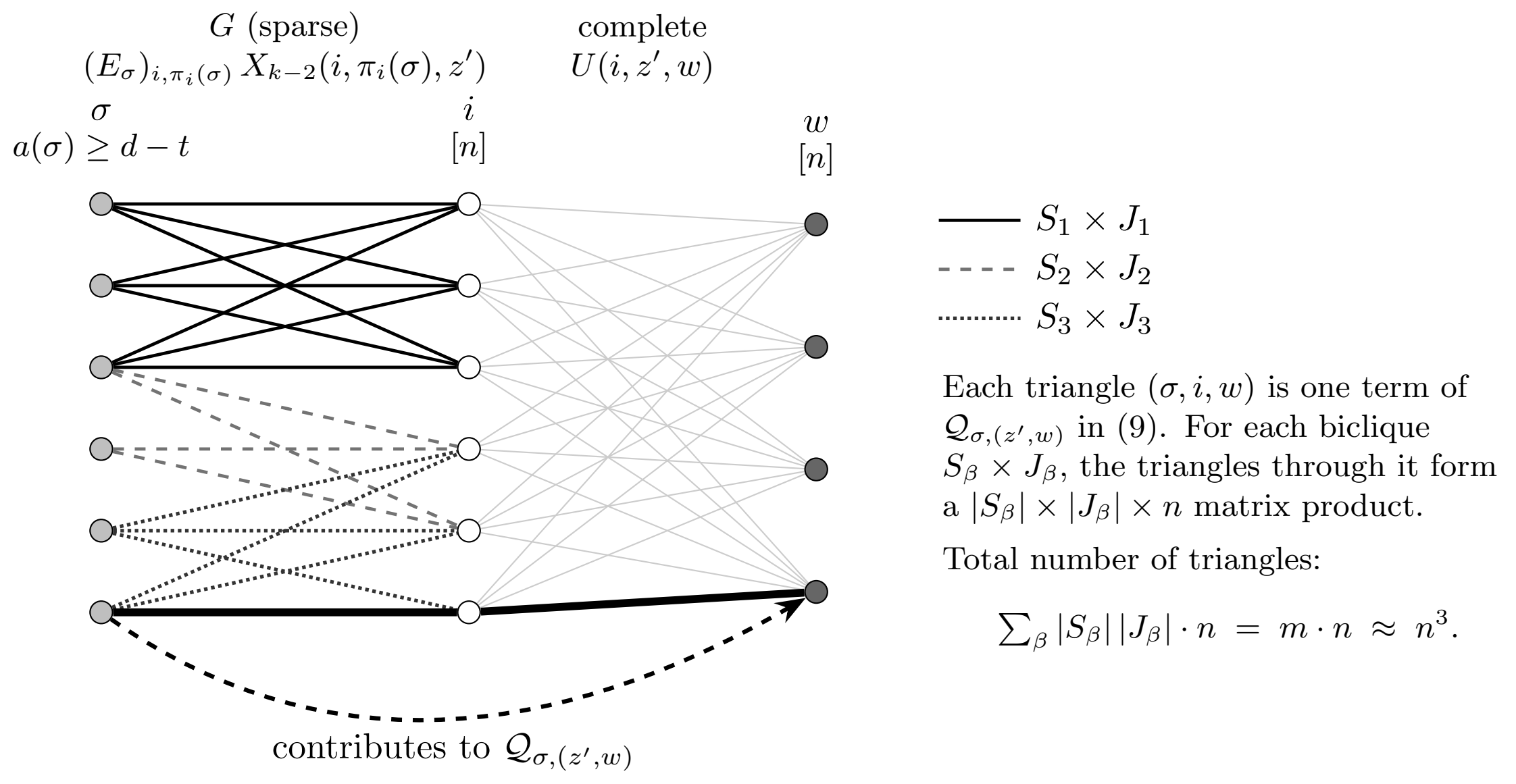

To formalize this partitioning, we consider the graph $G$ whose left vertices are the set of $\sigma$ with $a(\sigma) \ge d-t$, whose right vertices are $[n]$ (the rows of the $E_\sigma$ matrices), and where $\sigma \sim i$ when row $i$ of $E_\sigma$ is nonzero. In terms of this graph, we seek a partition of $E(G)$ into bicliques of size polynomial in $n$. This is achieved by \Cref{lem:biclique}. See \Cref{fig:Q} for an illustration.

To handle the $\calP$'s, we use a similar strategy. We wish to compute all (by symmetry of $E_\sigma$ we can use the same graph $G$ to represent nonzero columns)
\[\calP_{\sigma, z} = \sum_{j:(\sigma,j)\in E(G)} (E_\sigma)_{\pi_j(\sigma),j} V(j,z) \prod_{s=1}^{k-3} X_s(\pi_j(\sigma), j, z_{-s}).\]

First, note that after fixing $w$, this looks quite similar to a $(k-1)$-hyperclique polynomial where the $k-1$ coordinates are $\sigma, j$, and $z'$: each variable in each term omits exactly one coordinate --- the $V$ term is missing $\sigma$, and each $X_s$ variable omits one of the values of $z'$ (the $(E_\sigma)_{\pi_j(\sigma),j}$ can be absorbed into one of $X_s$). However, currently there are only $k-2$ variables in total, and none of them omits the value of $j$. This is not surprising, because the output $\calP_{\sigma, z}$ is exactly this missing ``term''. We can artificially multiply $\calP_{\sigma, z}$ by a new variable $Y(\sigma, z', w)$ and sum over all choices of $\sigma, z'$.
\[ \Phi_w \coloneqq \sum_{\sigma, z'} \calP_{\sigma, z} Y(\sigma, z', w) \]
This becomes the evaluation of a $(k-1)$-hyperclique polynomial. A classical result of Baur and Strassen~\cite{BaurStrassen83} tells us that there is an equivalence between evaluating $\Phi_w$ and all its derivatives. When one considers all derivatives on $Y$, this just recovers all the $\calP_{\sigma, z}$.

We will use our inductive hypothesis to construct a small circuit for $\Phi_w$, and we will invoke Baur--Strassen to compute \emph{all} such $\calP_{\sigma, z}$ with the same cost. As in the case of the $\calQ$'s, we run into the issue that the sum over $j$ depends on $\sigma$. We again use the partitioning of \Cref{lem:biclique} to fix this; the resulting circuit bound is our \Cref{lem:lowswap}.

To finish, we choose the degree threshold $t$ to be an appropriate linear function of $d$ so that both pieces of the circuit get a speedup. Adding the two pieces together gives the final circuit. Because the construction of the circuit is uniform and only uses constants from $\mathbb{Z}[1/2]$ (i.e., rational numbers of the form $a/2^b$ where $a$ is an integer and $b$ is a non-negative integer), our hyperclique counting algorithm \Cref{thm:countHC} follows immediately from evaluating this circuit.

\section{Preliminaries}

We will only need the following simple fact about fast matrix multiplication.

\begin{lemma}\label{lem:rect}
    Let $\ell = \min(n_1,n_2,n_3)$. There is an arithmetic circuit for $n_1 \times n_2$ by $n_2 \times n_3$ matrix multiplication of size $O((n_1n_2n_3) \cdot \ell^{-\gamma})$, for some constant $\gamma \ge 1/8$.
\end{lemma}

\begin{proof}
    By blocking our input matrices into $\ell \times \ell$ blocks, the matrix multiplication reduces to $n_1 n_2 n_3 / \ell^3$ instances of multiplying two $\ell \times \ell$ matrices. Each of those has an arithmetic circuit of size $O(\ell^{\log_2 7})$ by Strassen's algorithm. We may thus pick $\gamma = 3 - \log_2 7 > 1/8$. (State-of-the-art fast rectangular matrix multiplication could be used to achieve a better bound on $\gamma$, which is unnecessary for the main results here beyond slightly optimizing the exponents.)
\end{proof}

We will also use the following generalization of \Cref{lem:rect} to the hyperclique tensor. For a tensor $T$, let $\Ckt(T)$ denote the size of the smallest arithmetic circuit computing it.
Define the $n_1 \times n_2 \times \cdots \times n_k$ hyperclique tensor
\[\Delta^{(k)}_{(n_1, \ldots, n_k)} \coloneqq \sum_{\alpha \in [n_1] \times [n_2] \times \cdots \times [n_k]} ~ \prod_{j=1}^k x^{(j)}_{\alpha_{-j}}.\]
Write $\Delta^{(k)}_n$ to denote $\Delta^{(k)}_n \coloneqq \Delta^{(k)}_{\scriptscriptstyle \underbrace{\scriptscriptstyle(n,n,\ldots,n)}_{k \text{ times}}}$. 
\begin{lemma}\label{lem:tiling}
For $\ell \le \min_{i \in [k]} n_i$, the tensor $\Delta^{(k)}_{(n_1, \ldots, n_k)}$ has a circuit of size $\Ckt(\Delta^{(k)}_\ell) \cdot \prod_{i=1}^k \lceil n_i/\ell \rceil.$
\end{lemma}
\begin{proof}
Partition each index set $[n_i]$ into $n_i' \coloneqq \lceil n_i / \ell \rceil$ parts of size at most $\ell$, call them $P_{i,1}, \ldots, P_{i,n_i'}$. Then
\[\Delta^{(k)}_{(n_1, \ldots, n_k)} = \sum_{\substack{(i_1, \ldots, i_k) \in [n_1'] \times \cdots \times [n_k'] \\ J \in P_{1, i_1} \times \cdots \times P_{k,i_k}}} ~ \prod_{s=1}^k x^{(s)}_{J_{-s}}.\]
For each fixed choice of $i_1, \ldots, i_k$, the sum over $J$ can be computed by evaluating $\Delta^{(k)}_\ell$ (after possibly setting variables to zero). The claim follows.
\end{proof}

We will also make use of the Baur--Strassen theorem, which converts a circuit for computing a polynomial into one that computes its partial derivatives.

\begin{lemma}[\cite{BaurStrassen83,Morgenstern85}]\label{lem:Baur--Strassen}
Suppose a polynomial $P(x_1, \ldots, x_n)$ can be computed by an arithmetic circuit of size $s$. Then there exists an arithmetic circuit of size at most $4 \cdot s$ computing $\frac{\partial P}{\partial x_i}$ for all $i \in [n]$. Moreover, the new circuit can be constructed from the circuit for $P$ in time $O(s)$.
\end{lemma}

\section{Small arithmetic circuits for $\Delta^{(k)}_n$} \label{sec:mainproof}
In this section, we follow the overview  (\Cref{sec:overview}) to prove our main result via the following inductive step:
\begin{theorem} \label{thm:main}
Fix $k \ge 4$ and $\eps \in (0, 1/8]$. Suppose the $(k-1)$-hyperclique polynomial has a circuit of size $O(n^{k-1-\varepsilon})$ for all $n$.  Choose an integer $B \ge 16$ that satisfies $2^{\eps B} \ge (eB)^2.$
Then the circuit size of the $k$-hyperclique polynomial is $O(n^{k-\eps/(4B)})$.
\end{theorem}
We then appeal to the base case that $\Delta^{(3)}_n$ has circuits of size $n^{\omega+o(1)}$ (where $\omega < 3$ follows from any nontrivial fast matrix multiplication algorithm). Using the fact that all steps of the construction are uniform, we inductively obtain our main result:

\begin{reminder}{thm:hyperclique}
\end{reminder}

\subsection{Setup}
We begin by introducing parameters and recalling the setup of the proof overview. Throughout this section, fix $k$ and let $\varepsilon$ be the savings from the $k-1$ case. Choose an integer $B \ge 16$ that satisfies $2^{\eps B} \ge (eB)^2.$
Set
\[ q = 2^B, \qquad t = \lfloor d/B \rfloor. \]
We assume that $n$ is of the form $q^d$; this is innocuous, as we may pad to the next power of $q$ at a constant factor cost in the circuit size.
    
Label the $k$ coordinates as $(i,j,z_1, \ldots, z_{k-2}) \in [n]^k$. Introduce variables $U(i,z), V(j,z)$, and $X_s(i,j,z_{-s})$ for $i,j \in [n], z \in [n]^{k-2}, s \in [k-2]$. Then
\[\Delta^{(k)}_n = \sum_{i,j \in [n], z \in [n]^{k-2}} U(i, z) V(j,z) \prod_{s=1}^{k-2}X_s(i,j,z_{-s}). \]

Define matrices $Q^z$ and $P^z$ via
\[(Q^z)_{ij} = U(i,z) X_{k-2}(i,j,z_{-(k-2)}), \qquad (P^z)_{ij} = V(j,z) \prod_{s=1}^{k-3} X_{s}(i,j,z_{-s}).\]

We recall the transpose operators, which are instrumental to the decomposition of the identity matrix described in \Cref{sec:overview} above. View $q^d \times q^d$ matrices as having rows and columns indexed by vectors in $[q]^d$. For $p \in [d]$, the operator $\tau_p$ acts on the space of $q^d \times q^d$ matrices via

\[\tau_p \cdot Q_{(a_1, \ldots, a_p, \ldots,  a_d), (b_1, \ldots, b_p, \ldots, b_d)}  = Q_{(a_1, \ldots, b_p, \ldots,  a_d), (b_1, \ldots, a_p, \ldots, b_d)}.\]

Let $\calB$ be the basis for the space of $q \times q$ matrices given by
\[\mathcal{B} = \{e_{ii}\}_{i \in [q]} \cup \{e_{ij} + e_{ji}, e_{ij} - e_{ji}\}_{1 \le i < j \le q}.\] 

Then $\calB^{\otimes d}$ is a basis for the space of $n \times n$ matrices. For $\sigma = (\sigma_1, \ldots, \sigma_d)$ with $\sigma_i \in \mathcal{B}$, let $E_\sigma = \bigotimes_{i \in [d]} \sigma_i$. Let $\pi_i(\sigma)$ denote the column index of a nonzero entry in row $i$ in $E_\sigma$. Since the support of $E_\sigma$ is symmetric, $\pi_j(\sigma)$ also denotes the row index of the nonzero entry in column $j$. Let $u(\sigma)$ be the number of non-diagonal factors in $\sigma$, and let $a(\sigma)$ be the number of skew-symmetric factors in $\sigma$ (so $a(\sigma) \le u(\sigma))$.

The following are immediate:
\begin{lemma}\label{lem:basis}
    \begin{enumerate}
    \item $\langle E_\sigma, E_{\sigma'} \rangle = 0$ if $\sigma \neq \sigma'$, and $2^{u(\sigma)}$ otherwise.
        \item $\tau_p E_\sigma = E_\sigma$ if $\sigma_p$ is diagonal or symmetric, and $\tau_p E_\sigma = -E_\sigma$ if $\sigma_p$ is skew-symmetric.
        \item $(I-\tau_p)E_\sigma = 0$ if $\sigma_p$ is symmetric or diagonal, and $(I-\tau_p)E_\sigma = 2E_\sigma$ if $\sigma_p$ is skew-symmetric.
    \end{enumerate}
\end{lemma}

Define the low and high degree operators
\[ L = \sum_{|A| \le t} \left(\prod_{a \in A} \tau_a\right) \left(\prod_{a \notin A} (I-\tau_a)\right), \qquad 
H = \sum_{|A| > t} \left(\prod_{a \in A} \tau_a\right) \left(\prod_{a \notin A} (I-\tau_a)\right)
\]
recalling that $t = \lfloor d/B \rfloor$.

For a linear map $T$ on the space of $n \times n$ matrices, define
\[\Delta^{(k)}_{n,T} \coloneqq \sum_{z} \langle P^{z}, T \cdot Q^{z} \rangle.\]
Then $\Delta^{(k)}_{n} = \Delta^{(k)}_{n,L} + \Delta^{(k)}_{n,H}$.

We first give a circuit for $\Delta^{(k)}_{n,H}$ (\Cref{cor:finalhigh} in \Cref{subsec:high}), then give a circuit for $\Delta^{(k)}_{n,L}$ (\Cref{lem:lowswap} in \Cref{subsec:low}), and finally combine them (proving \Cref{thm:main} in \Cref{subsec:final}).

\subsection{High-degree}\label{subsec:high}
We begin by looking at the result of applying a particular partial transpose $\tau_A$ with a large number of swaps to our computation.
\begin{lemma}\label{lem:highswap}
$\Delta^{(k)}_{n,\tau_A}$ has circuits of size $O(n^k q^{- \varepsilon |A|})$.
\end{lemma}
\begin{proof}
To make the effects of the partial transpose a bit clearer, we introduce the following notation: write $i = (i_A, i_{\overline{A}}),~ j = (j_A, j_{\overline{A}})$ and $i^* = (j_A, i_{\overline{A}}),~ j^* = (i_A, j_{\overline{A}})$. Then by definition, \[(\tau_A \cdot Q)_{(i_A, i_{\overline{A}}),(j_A, j_{\overline{A}})} = Q_{(j_A, i_{\overline{A}}),(i_A, j_{\overline{A}})} = Q_{i^*, j^*}.\]
By definition of $\Delta^{(k)}_{n,\tau_A}, P^z,$ and $Q^z$, we have
\begin{align*}
\Delta^{(k)}_{n,\tau_A} &= \sum_z \langle P^{z}, \tau_A \cdot Q^{z} \rangle = \sum_z \sum_{i,j} (P^z)_{ij} (Q^z)_{i^*,j^*} \\
&= \sum_{z,i,j} \left(V_{j,z} \prod_{s=1}^{k-3} X_{s}(i,j,z_{-s})\right) (U_{i^*,z}X_{k-2}(i^*,j^*,z_{-(k-2)}))\\
&= \sum_{j_A, j_{\overline{A}}, i_{\overline{A}}} \sum_{z,i_A} (V_{j,z}U_{j_A, i_{\overline{A}},z}) \cdot X_{k-2}(i^*, j^*,z_{-(k-2)})\cdot \prod_{s=1}^{k-3} X_{s}(i,j,z_{-s}).
\end{align*}
The key point is that the partial transpose replaces $U$'s dependence on $i_A$ with a dependence on $j_A$. After we fix $j$ and $i_{\overline{A}}$, both $U$ and $V$ only depend on $z$, and we can combine them into one input. This reduces the number of inputs from $k$ to $k-1$, allowing us to invoke the inductive hypothesis.

For each choice of $j_A, j_{\overline{A}}, i_{\overline{A}}$, of which there are $n^2/q^{|A|}$, the inner sum is an imbalanced $(k-1)$-hyperclique tensor with indices ranging over $[q^{|A|}] \times [n] \times \cdots \times [n]$ evaluated on inputs 
\[(V_{j,z}U_{j_A, i_{\overline{A}},z})_{z}, \quad (X_{k-2}(i^*,j^*,z_{-(k-2)}))_{i_A, z_{-(k-2)}}, \quad (X_s(i,j,z_{-s}))_{i_A, z_{-s}}, s \in [k-3].\]
By \Cref{lem:tiling}, and using $n^{k-2}$ multiplications to compute $(V_{j,z}U_{j_A, i_{\overline{A}},z})$ for all $z \in [n]^{k-2}$, this tensor has a circuit of size 
\[(n/q^{|A|})^{k-2} \cdot q^{|A|(k-1-\varepsilon)} + n^{k-2} = O(n^{k-2} q^{(1-\varepsilon) |A|}).\]
Summing over all $j,i_{\overline{A}}$ yields a circuit of the claimed size.
\end{proof}

Thus, whenever $|A|$ is such that $q^{-\eps |A|}$ is polynomial in $n,$ (i.e. the number of swaps is sufficiently large), we achieve nontrivial savings. The following corollary formalizes this intuition:
\begin{corollary}\label{cor:finalhigh}
    $\Delta^{(k)}_{n,H}$ has circuits of size $O(n^{k - \varepsilon/(2B)})$.
\end{corollary}
\begin{proof}
We have that
\[\Delta^{(k)}_{n,H} = \sum_{|A| > \lfloor d/B \rfloor} c_A \, \Delta_{n,\tau_A}^{(k)},\]
where the integers $c_A$ (with $|c_A| \le 2^{|A|}$) come from expanding the products $\prod_{a \notin A}(I - \tau_a)$ in the definition of $H$.
Applying \Cref{lem:highswap} to compute each term gives a circuit for $\Delta^{(k)}_{n,H}$ of size
\[n^k \sum_{a > \lfloor d/B \rfloor} \binom{d}{a} \cdot 2^{-B\varepsilon a}.\]
We have $\binom{d}{a} \leq \left(\frac{ed}{a}\right)^a < (eB)^a$. Because we took $B$ to satisfy $2^{\eps B} \ge (eB)^2$, $\binom{d}{a} \le 2^{\eps B a/2},$ so this is at most
\[
n^k\sum_{a > \lfloor d/B \rfloor}^d 2^{-\eps B a/2} \le n^k \cdot 2^{-\eps d/2}  \cdot \frac{1}{1-2^{-\eps B/2}} \le n^k \cdot 2^{-\varepsilon d/2} \cdot \frac{1}{1-(eB)^{-1}} \le O(n^k \cdot 2^{-\eps d/2}).\]
The first inequality comes from the fact that the sum is a geometric series in $a$ with initial term $2^{- \eps B(\lfloor d/B \rfloor + 1)/2} < 2^{-\varepsilon d/2}$ and ratio $2^{-\varepsilon B/2}$. The second follows by the aforementioned bound on $B$. Finally, since $n = 2^{Bd}$, this is at most $O(n^{k-\eps/(2B)})$.
\end{proof}
\subsection{Low-degree}\label{subsec:low}
Throughout this section, we will consider the following graph to understand (and eventually, block) the contribution of the low-degree part. Let $G$ be the bipartite graph whose left vertices are the labels $\sigma \in \calB^{\otimes d}$ such that the number of antisymmetric factors, $a(\sigma)$, is at least $d-t = d- \floor{d/B}$. The right vertices are the rows $i \in [n],$ and the edges are pairs $(\sigma,i)$ such that the $i$-th row of $E_\sigma$ is nonzero. By symmetry, the graph also denotes all pairs $(\sigma, j)$ such that the $j$-th column is nonzero. Let $m = |E(G)|$.

\begin{lemma} \label{lem:edge_bound}
    The number of edges in $G$ is at most $O(n^{2+\eps/(2B)}).$
\end{lemma}
\begin{proof}
    Each basis element $E_\sigma$ has $2^{u(\sigma)}$ nonzero columns, and therefore, each left vertex has $2^{u(\sigma)}$ incident edges. So the total number of edges is given by $\sum_{\sigma : a(\sigma) \ge d-t} 2^{u(\sigma)}$. Consider the contribution to this sum for all $\sigma$ with $a(\sigma) = \alpha$. Fix a choice of antisymmetric positions, of which there are $\binom{d}{\alpha}$. Then there are $\binom{q}{2}$ basis elements to choose from, and each such position contributes a factor of $2$ to the number of outgoing edges. Similarly, for each remaining position, we can choose $\binom{q}{2}$ symmetric basis elements or $q$ diagonal basis elements, and each symmetric position contributes a factor of $2$ to the number of outgoing edges. Thus, the total contribution for a fixed set of antisymmetric positions equals \[\left[2\binom{q}{2} \right]^\alpha \cdot \left[2\binom{q}{2}  + q\right]^{d-\alpha} \le q^{2d} = n^2.\]
     Therefore, the number of edges in $G$ is at most
     \[
         n^2 \sum_{\alpha = d-t}^d \binom{d}{\alpha} \le n^2 \sum_{i=0}^t \binom{d}{i} \le n^2 B^{d/B} \sum_{i=0}^d \binom{d}{i} B^{-i} \le n^2 B^{d/B}\left(1+\frac{1}{B}\right)^d \le n^2(eB)^{d/B}.
     \]
     The first inequality results from taking $i = d-\alpha$ for all $\alpha.$ The second is because $t \le d/B$ and therefore $B^{d/B - i} \ge 1$ for $i \le t$. The third one uses the binomial theorem, and the final one uses the fact that $1+x \le e^x,$ taking $x=1/B.$ We can now apply our assumption that $(eB)^2 \le 2^{\eps B},$ as well as the fact that $n = 2^{Bd}$, which gives
     \[ |E(G)| \le n^2 2^{\eps d/2} \le n^{2+\eps/(2B)}, \]
    as desired.
\end{proof}

\begin{lemma} \label{lem:biclique}
The edge set $E(G)$ can be partitioned into a disjoint union of complete bipartite subgraphs $S \times J$ such that $|S|, |J| \ge \ell,$ where $\ell \coloneqq 2^{(B-1) \floor{(d-t)/B}}$. 
\end{lemma}
\begin{proof}
    The edges in question are of the form $(\sigma, j)$. Recall that $\sigma \in \mathcal{B}^d$, and that while $j$ is an index in $[n],$ we can think of it as being an element of $[q]^d$. We say that the \emph{type} of $\sigma$ is the $d$-tuple that details whether each of the basis vectors $\sigma_1, \dots, \sigma_d$ are diagonal, symmetric, or antisymmetric. Fix a choice of type. Upfront, we will arbitrarily partition the off-diagonal positions into two categories, which we will call $\mathsf{Column}$ and $\mathsf{Basis}$, such that  $|\mathsf{Column}| \eqqcolon \rho.$ Let $\mathsf{Diag}$ be the set of diagonal positions. We will characterize a biclique in the following way: for each $p \in \mathsf{Column} \cup \mathsf{Diag},$ choose $c_p \in [q]$ arbitrarily. For each $p \in \mathsf{Basis},$ choose a basis element $b_p \in \mathcal{B}$ whose type matches that of $p$. We will add an edge $(\sigma, j)$ to this particular biclique if for every position $p$, the following hold:
    \begin{itemize}
        \item if $p \in \mathsf{Column},$ we require that the column support of $\sigma_p$ contains $c_p$ and $j_p=c_p$. There are $q-1$ choices for $\sigma_p$, one for each possible other endpoint of its support.
        \item if $p \in \mathsf{Basis},$ we require that $\sigma_p = b_p$ and that $j_p$ is in the support of $b_p.$ Thus, if $b_p = e_{xy} \pm e_{yx},$ there are two valid choices of $j_p$, which are $x$ and $y.$
        \item if $p \in \mathsf{Diag},$ we require that $\sigma_p=e_{c_pc_p}$ and $j_p=c_p$.
    \end{itemize}
    By the definition, each $(\sigma, j)$ maps to exactly one of these bicliques. For every $\sigma\in S$ and $j\in J$, our restrictions enforce that $j_p$ is in the support of $\sigma_p$ at every position, which is why every edge in $S \times J$ must exist.

    All that remains is to bound the sizes of $S$ and $J$. This follows from the number of elements that can satisfy each of our restrictions. In particular, we have that $|S| = (q-1)^\rho$ and $|J| = 2^{u-\rho},$ where $u$ is the number of off-diagonal positions. We have that $q = 2^B$, meaning $q-1 \ge 2^{B-1}.$ Thus, $|S| \ge 2^{(B-1)\rho}.$ Note that the number of antisymmetric positions is at least $d-t$.  Thus, we have that $|J| = 2^{u-\rho} \ge 2^{d-t-\rho}$. Balancing these gives $\rho = \floor{(d-t)/B},$ and substituting these into the lower bounds on $|S|$ and $|J|$ completes the proof.
\end{proof}

\begin{lemma}\label{lem:lowswap}
$\Delta^{(k)}_{n,L}$ has circuits of size $O(n^{k - \eps/(4B)})$.
\end{lemma}
\begin{proof}

By \Cref{lem:basis}, the matrices $E_\sigma$ form an orthogonal
basis with $\langle E_\sigma,E_\sigma\rangle = 2^{u(\sigma)}$.
Each $E_\sigma$ is an eigenvector of every $\tau_p$,
and hence also of $L$. We first determine its eigenvalue.

Reindexing the definition of $L$ by replacing each indexing set
with its complement gives
\[
L =
\sum_{\substack{C\subseteq[d]\\ |C|\ge d-t}}
\left(\prod_{p\in C}(I-\tau_p)\right)
\left(\prod_{p\notin C}\tau_p\right).
\]
Let $A_\sigma\subseteq[d]$ be the set of antisymmetric positions
of $E_\sigma$, so that $|A_\sigma|=a(\sigma)$.
By \Cref{lem:basis}, the summand indexed by $C$ annihilates
$E_\sigma$ unless $C\subseteq A_\sigma$.
If $C\subseteq A_\sigma$ and $|C|=c$, each position in $C$
contributes a factor of $2$, each antisymmetric position outside
$C$ contributes a factor of $-1$, and every remaining position
contributes a factor of $1$. Thus
\[
LE_\sigma=\lambda_\sigma E_\sigma,
\qquad
\lambda_\sigma =
\sum_{c=d-t}^{a(\sigma)}
\binom{a(\sigma)}{c}\,2^c(-1)^{a(\sigma)-c}.
\]
In particular,
$\lambda_\sigma=0$ whenever $a(\sigma)<d-t$. Expanding $Q^z$ in the orthogonal basis gives us:
\[
Q^z =
\sum_\sigma
\frac{\langle Q^z,E_\sigma\rangle}{2^{u(\sigma)}}E_\sigma.
\]
We can now apply $L$ and take the inner product with $P^z$, as follows:
\[
\Delta^{(k)}_{n,L}
= \sum_z \langle P^z,LQ^z\rangle
= \sum_{\substack{z,\sigma\\a(\sigma)\ge d-t}}
\frac{\lambda_\sigma}{2^{u(\sigma)}}
\langle P^z,E_\sigma\rangle
\langle Q^z,E_\sigma\rangle.
\]

    We will bound this sum using the properties of the graph mentioned above. Let \begin{equation}\label{eq:cliquedecomp}
        E(G) = \bigsqcup_\beta (S_\beta \times J_\beta)
    \end{equation}
    be the decomposition into bicliques of \Cref{lem:biclique}.

    We begin by computing the values of $\calQ_{\sigma, z} = \ip{E_\sigma, Q^z}$. Write $z = (z', w),$ where $z'$ is a $(k-3)$-tuple of indices. We have
    \begin{align*}
        \calQ_{\sigma, (z', w)} = \sum_{i: (\sigma, i) \in E(G)} (E_\sigma)_{i, \pi_i(\sigma)} X_{k-2}(i, \pi_i(\sigma), z') U(i,z',w).
    \end{align*}
    We need to compute this quantity for all $\sigma$ and $z$. We will use \Cref{lem:biclique} to partition this computation.
    
    Fix a biclique $S_\beta \times J_\beta$ from \Cref{eq:cliquedecomp} and a choice of $z'.$ Let $A^\beta$ be the $|S_\beta| \times |J_\beta|$ matrix with
    \[A^\beta_{\sigma, i} = (E_\sigma)_{i, \pi_i(\sigma)} X_{k-2}(i, \pi_i(\sigma), z')\]
    and let $B^\beta$ be the $|J_\beta| \times n$ matrix with
    \[B^\beta_{i,w} = U(i,z',w).\]
    Then 
    \[(A^\beta B^\beta)_{\sigma, w} = \sum_{i \in J_\beta } (E_\sigma)_{i, \pi_i(\sigma)} X_{k-2}(i, \pi_i(\sigma), z') U(i,z',w)\]
    is the restriction of the sum $\calQ_{\sigma, (z', w)}$ to $J_\beta$. Hence $\calQ_{\sigma, (z',w)} = \sum_\beta (A^\beta B^\beta)_{\sigma,w}$, where the sum is over the $\beta$ with $\sigma \in S_\beta$. Thus it suffices to compute all products $A^\beta B^\beta$ to compute all $\calQ$'s.
    
    From \Cref{lem:biclique}, $|S_\beta|, |J_\beta| \ge \ell$. It follows from \Cref{lem:rect} that the product $A^\beta B^\beta$ can be computed with $O(|S_\beta||J_\beta|n /\ell^\gamma)$ operations for some $\gamma > 1/8$. The cost of computing these products over all bicliques and all $z'$ equals 

    \[ n^{k-3} \sum_\beta |S_\beta||J_\beta|n /\ell^\gamma = O(n^{k-2} m /\ell^\gamma) \] because each of the $m$ edges in $G$ is covered by exactly one biclique, and there are $|S_\beta| \cdot |J_\beta|$ edges per biclique. Finally, since $\gamma >1/8 \ge \varepsilon$, this is upper bounded by $O(n^{k-2} m /\ell^\varepsilon)$.

    We now turn to computing the values of $\calP_{\sigma, z} = \ip{E_\sigma, P^z}$, which we can write as 
    \[\calP_{\sigma, z} = \sum_{j:(\sigma,j)\in E(G)} (E_\sigma)_{\pi_j(\sigma), j} V(j,z) \prod_{s=1}^{k-3} X_s(\pi_j(\sigma), j, z_{-s}). \]

    Fix a biclique $\beta$ and a choice of $w \in [n]$. Introduce new variables $Y(\sigma, z',w)$ for $\sigma \in S_\beta$ and $z' \in [n]^{k-3}$. Let $\Phi = \Phi_{\beta,w}$ be the polynomial \[ \Phi_{\beta,w} = \sum_{\sigma \in S_\beta,\, j \in J_\beta,\, z' \in [n]^{k-3}} (E_\sigma)_{\pi_j(\sigma), j}\, Y(\sigma, z',w)\, V(j,(z',w)) \prod_{s=1}^{k-3} X_s(\pi_j(\sigma), j, (z',w)_{-s}). \]

    This is an imbalanced $(k-1)$-hyperclique tensor after scaling inputs (we may scale an $X_s$ variable by $(E_\sigma)_{\pi_j(\sigma), j}$, since the result is still a function of $j$ and $\sigma$) of shape $|S_\beta| \times |J_\beta| \times n \times \cdots \times n$. Recall that $\ell = 2^{(B-1) \lfloor (d-t)/B \rfloor} \leq 2^{d-t} \leq n$ since $t \le d$, and also $|S_\beta|, |J_\beta| \ge \ell$ by the guarantee of \Cref{lem:biclique}. Hence by \Cref{lem:tiling} and our inductive assumption, $\Phi$ can be computed by a circuit of size 
    \[\ell^{k-1-\varepsilon} \cdot \frac{n^{k-3}|S_\beta||J_\beta|}{\ell^{k-1}} = n^{k-3}|S_\beta||J_\beta| \cdot \ell^{-\varepsilon}.\]
    By \Cref{lem:Baur--Strassen}, we can compute the partial derivatives of $\Phi$ w.r.t.~all $Y$ variables with a circuit of the same size. Note that
    $\calP_{\sigma, (z',w)} = \sum_{\beta : \sigma \in S_\beta} \frac{\partial \Phi_{\beta,w}}{\partial Y(\sigma, z',w)}$, since the sets $J_\beta$ with $\sigma \in S_\beta$ partition the neighbors of $\sigma$ in $G$. Hence we can compute all $\calP_{\sigma, z}$ from the polynomials $\Phi_{\beta,w}$ for all $w$ and $\beta$. This gives a circuit for all $\calP$'s of size
    \[\sum_{w, \beta} n^{k-3}|S_\beta||J_\beta| \cdot \ell^{-\varepsilon} = n^{k-2} \ell^{-\varepsilon} \sum_\beta |S_\beta| |J_\beta| = n^{k-2} m \ell^{-\varepsilon}.\]
        Finally, given all $\calP_{\sigma,z}$ and $\calQ_{\sigma,z}$, we compute
    \[\Delta^{(k)}_{n,L} = \sum_{z,\ \sigma : a(\sigma) \ge d-t} \frac{\lambda_\sigma}{2^{u(\sigma)}}\, \calP_{\sigma,z}\, \calQ_{\sigma,z}\]
    using $O(n^{k-2} \cdot \#\sigma)$ operations, where $\#\sigma$ is the number of left vertices of $G$. Each left vertex has $2^{u(\sigma)} \ge 2^{a(\sigma)} \ge 2^{d-t}$ incident edges, so $\#\sigma \le m/2^{d-t} \le m/\ell$. We form the entries of the matrices $A^\beta$, and the inputs of the polynomials $\Phi$ for all $\beta$ and $w$, using $O(n^{k-3} m)$ operations. Since $\ell \le n$, both of these operation counts are $O(n^{k-2} m/\ell^{\varepsilon})$.

    Finally, we need to bound $\ell,$ the lower bound on the size of the bicliques. We have that
    \begin{align*}
        \log \ell \coloneqq (B-1)\left\lfloor\frac{d-t}{B}\right\rfloor &\ge \frac{d(B-1)^2}{B^2}-(B-1),
        \intertext{which results from using $t \le d/B$ and removing the floor function. Since $n=2^{Bd},$ we have $d=\log n /B,$}
        \log \ell \ge \frac{(B-1)^2}{B^3}\log n - (B-1) &= \frac{(1-1/B)^2}{B}\log n - (B-1),
    \end{align*}
    and exponentiating tells us that $\ell \ge 2^{-(B-1)} n^{(1-1/B)^2/B}.$ $B$ is independent of $n,$ and because $B\ge 16,$ we have that $(1-1/B)^2 \ge 3/4.$ So we have $\ell \ge \Omega(n^{3/(4B)}).$

    Applying this along with \Cref{lem:edge_bound} tells us that the overall circuit size is, up to constants, at most  
    \[ \frac{n^{k-2} m}{\ell^{\varepsilon}} \le \frac{n^{k-2}\cdot n^{2+\eps/(2B)}}{n^{3\eps /(4B)}} \le n^{k - \eps/(4B)}. \qedhere \]
\end{proof}

\subsection{The final circuit}\label{subsec:final}

By combining \Cref{cor:finalhigh} (the high degree part) and \Cref{lem:lowswap} (the low degree part), we are finally ready to prove our main theorem. 
\begin{proof}[Proof of \Cref{thm:main}]
    Recall that
    \[ \Delta^{(k)}_n = \Delta^{(k)}_{n,L} + \Delta^{(k)}_{n,H}. \]

    By \Cref{cor:finalhigh}, $\Delta^{(k)}_{n,H}$ has a circuit of size $O(n^{k-\varepsilon/(2B)})$. By \Cref{lem:lowswap}, $\Delta^{(k)}_{n,L}$ has a circuit of size $O(n^{k-\varepsilon/(4B)})$. Adding these circuits together gives the desired circuit of size $O(n^{k-\varepsilon/(4B)})$. 
\end{proof}

While the above argument suffices to show that we can attain \emph{some} nontrivial savings for every value of $k$, we would like to look at the way that the savings shrink as $k$ increases. Since we lose a factor of $1/(4B)$ at each step of the induction, it is clear that our advantage deteriorates at least exponentially in $k.$ It turns out that because we require that $2^{\eps B} \ge (e B)^2$ at every iteration, we need to define new values of $B$ at each stage too. Ultimately, we get a doubly exponential decay in the savings, as we will show next, which completes the proof of the main theorem.

\begin{proof}[Proof of \Cref{thm:hyperclique}] 
    Let $\eps_j$ be the savings for the $j$-hyperclique polynomial. Already from Strassen's algorithm, we have that $\eps_3 = 1/8.$ While it is possible to optimize these constants further, the asymptotics do not change.
    We will define \[L_j \coloneqq \log(1/\eps_j) \qquad B_j \coloneqq \left\lceil \frac{8L_j}{\eps_j} \right\rceil 
    \qquad \eps_{j+1} \coloneqq\frac{\eps_j}{4B_j}
    \]
    We first need to check that $2^{\eps_j B_j} \ge (e B_j)^2$ for all $j.$ Since $B_j \le 9L_j/\eps_j$ and $L_j \ge 3$ (as $\eps_j \le 1/8)$, we have  
    \[ \log(eB_j) \le \log(9e) + \log(L_j) - \log(\eps_j) \le  5 + \log(L_j) + L_j \le 4L_j \le \eps_j B_j / 2. \] Doubling and exponentiating this proves the inequality. We will show that $L_j \le 2^{j+1}-j-8$ via induction. Since $L_3 \le 3,$ the base case holds.
    
    By the upper bound on $B_j,$ we have $\eps_{j+1} \ge \eps^2_j/(36L_j).$ Then
    \[ L_{j+1} = -\log(\eps_{j+1}) \le -\log(\eps_j^2) + \log(36L_j) \le 2L_j + \log L_j + 6. \] 
    Using the inductive hypothesis gives $L_{j+1} \le 2^{j+2} - (j+1) - 8$, as desired. So $L_k = O(2^{k}),$ and we have that $\eps_k = 2^{-L_k},$ which makes the circuit size at most $O\left(n^{k-2^{-O(2^k)}}\right)$.

    Finally, we briefly remark on the uniformity. All the constants occurring in the circuit can be efficiently constructed and are not the bottleneck. Each step of our circuit construction is evidently efficient. The only technical input is Baur--Strassen (\Cref{lem:Baur--Strassen}) which is also known to be implementable in linear time (see~\cite{Morgenstern85} for the proof).
\end{proof}

\section{Applications}

We have shown (\Cref{thm:hyperclique}) that the  $k$-hyperclique polynomial $\Delta^{(k)}_n$ has uniformly constructible circuits of size $O(n^{k-\eps_k})$, where $\eps_k = 2^{-\Theta(2^k)}$. Moreover, all constants in the circuit are in ${\mathbb Z}[1/2]$, so we can evaluate it faithfully over any commutative ring in which $2$ is a unit, such as ${\mathbb Z}/(M {\mathbb Z})$ for odd $M$, or any finite field with odd characteristic.
We now give several applications of this result.

\def \F {{\mathbb F}}

\subsection{Counting Hypercliques}

\begin{reminder}{thm:countHC} 
\end{reminder}

\begin{proof} Make $k$ copies $V_1,\ldots,V_k$ of the vertices of $H$. For all $i \in [k]$ and $\alpha \in \prod_{j \neq i} V_j$, we make an assignment $a$ to the variables of $\Delta^{(k)}_n$: set $a^{(i)}_{\alpha} = 1$ if $\alpha$ forms a hyperedge in $H$, otherwise $a^{(i)}_{\alpha} = 0$. Observe that $\Delta^{(k)}_n(a)$ equals the number of $k$-tuples of vertices that form a $k$-hyperclique; this is precisely $k!$ times the number of hypercliques. By \Cref{thm:hyperclique}, we can construct the circuit for $\Delta^{(k)}_n$ in $O(n^{k - \eps_k+o(1)})$ time. Note the value of $\Delta^{(k)}_n(a)$ is at most $n^k$. To preserve its value, we can deterministically evaluate the circuit for $\Delta^{(k)}_n(a)$ modulo $M$ for any odd $M > n^k$, in $n^{k - \eps_k+o(1)}$ time.
\end{proof}

Similarly, we can also obtain simple applications to hypergraph algorithms by extending the known applications of matrix multiplication. For example, we can compute the number of $4$-hypercliques in a $3$-uniform hypergraph on $m$ hyperedges in $m^{4/3-\eps}$ time for some $\eps > 0$ (by extending the high-degree/low-degree technique of \cite{AYZ97} on triangles), we can find maximum node-weight hypercliques faster (extending the method of \cite{DBLP:journals/siamcomp/CzumajL09} for triangles), and we can similarly list hypercliques faster (extending \cite{DBLP:conf/icalp/MathialaganWX22} and other works). We omit the details.

\subsection{k-wise matrix product}

Next, we show how to compute the $k$-dimensional generalization of matrix multiplication, the \emph{$k$-wise matrix product}~\cite[Section~5.1]{BS21}. Given $k$ tensors $A_1,\dots,A_k$ of order $k$ with dimensions
$n\times\dots\times n$, this is defined as:
\[\mathrm{MP}_k(A_1,\dots,A_k)[i_1,\dots,i_k]=\sum_{\ell=1}^{n}A_1[\ell,i_2,\dots,i_k]\,A_2[i_1,\ell,i_3,\dots,i_k]\cdots A_k[i_1,\dots,i_{k-1},\ell].\]

\begin{reminder}{thm:k-dim-prod}
\end{reminder}

The basic idea is that we may view $\mathrm{MP}_k$ as the \emph{gradient} of the $(k+1)$-hyperclique polynomial $\Delta^{(k+1)}_n$, so we can apply Baur--Strassen~(\Cref{lem:Baur--Strassen}) to our circuit for $\Delta^{(k+1)}_n$ to compute $\mathrm{MP}_k$.

\begin{proof} Start by indexing the $k+1$ coordinates of $\Delta^{(k+1)}_n(X)$ by $(i_1,\ldots,i_k,\ell)$. For every $j \in [k]$, the group of inputs omitting coordinate $i_j$ is indexed by the $k$-tuples 
\[(i_1,\ldots,i_{j-1},i_{j+1},\ldots,i_k,\ell),\] 
and the group omitting $\ell$ is indexed by $(i_1,\ldots,i_k)$. For all $j \in [k]$, set
\[x^{(j)}_{(i_1,\ldots,i_{j-1},i_{j+1},\ldots,i_k,\ell)} \coloneqq A_j[i_1,\ldots,i_{j-1},\ell,i_{j+1},\ldots,i_k].\]
That is, the index $\ell$ goes in position $j$ of $A_j$. Leave the $x^{(k+1)}_{\alpha}$ variables free. Letting $\alpha \in [n]^k$, the monomial of $(\alpha,\ell) = (i_1,\ldots,i_k,\ell)$ becomes 
\[A_1[\ell,i_2,\ldots,i_k] \cdots A_k[i_1,\ldots,i_{k-1},\ell]\cdot x^{(k+1)}_{\alpha}.\] Summing over $\ell$, we have 
\[\Delta^{(k+1)}_n(A_1,\ldots,A_k,x^{(k+1)}) 
= \sum_{\alpha \in [n]^k} \mathrm{MP}_k(A_1,\ldots,A_k)[\alpha] \cdot x^{(k+1)}_{\alpha}.\]
Observe that this polynomial is linear in that last group of variables, and its partial derivative with respect to each $x^{(k+1)}_{\alpha}$ is the entry $\alpha \in [n]^k$ of the desired product. 

By the Baur--Strassen theorem (\Cref{lem:Baur--Strassen}), all partial derivatives of the above polynomial can be computed by an arithmetic circuit of size that is linear in the circuit size of the $\Delta^{(k+1)}_n$ polynomial; moreover, we can build the partial derivative circuit from the original one in linear time. Since our original circuit has size $O(n^{k+1-\eps_{k+1}})$, we obtain an efficient circuit for $\mathrm{MP}_k$.

Finally, to evaluate this circuit over Boolean inputs, we note that each output is a sum of $n$ products of inputs, so it is an integer between $0$ and $n$. As in the proof of \Cref{thm:countHC}, we may work modulo a prime larger than $n$ and recover the output exactly. 
\end{proof}

\subsection{Max-kSAT, Max-kCSP, etc}

Here, we review why the \MaxkSat hypothesis (Hypothesis~\ref{hyp:maxksat}) is falsified. Recall that we claim to solve the more general {\sf Max-Degree-$k$CSP} problem. A Boolean function on $n$ variables is a \emph{degree-$k$ constraint} if it has degree at most $k$ in its unique multilinear polynomial representation over the $\{0,1\}$ basis. (Note that \MaxkSat represents the very special case where each polynomial is over at most $k$ variables.) The {\sf Max-Degree-$k$CSP} problem is: given any set of degree-$k$ constraints over $n$ Boolean variables, find a Boolean assignment that sets a maximum number of constraints to true.

\begin{reminder}{thm:maxksat} 
\end{reminder}

\begin{proof}
    Theorem~C.1 in \cite{lincoln2018tight} shows that, for all constants $\ell,k$, if $\ell$-hypercliques in a $k$-uniform hypergraph with $N$ vertices can be detected in $T(N)$ time, then {\sf Max-Degree-$k$CSP} can be solved in time
    \begin{align}\label{eq:degk}
    \tilde O\left((mn^k)^{\binom{\ell}{k}}\cdot T(\ell \cdot 2^{n/\ell}) + mn^k \cdot 2^{kn/\ell}\right).
    \end{align}
    (The special case of \MaxkSat is covered by their Lemma 9.1.) Set $\ell = k+1$, so $\binom{\ell}{k} = k+1$. Applying \Cref{thm:countHC} with $k+1$ in place of $k$, we have $T(N) \leq O(N^{k+1 - \eps_{k+1}})$. For fixed $k$, the expression \eqref{eq:degk} is then
    \[\poly(m,n) \cdot (2^{n(k+1 - \eps_{k+1})/(k+1)} + 2^{kn/(k+1)}) \leq 2^{n(1-\delta_k)} \cdot \poly(m,n)\] for $\delta_k \coloneqq \eps_{k+1}/(k+1)$. This completes the proof.
\end{proof}

Inspecting their proof (which generalizes the algorithm for {\sf Max-Degree-$2$CSP} in \cite{Williams07thesis}), we may also exactly count the number of Boolean assignments that maximize the number of satisfied constraints in the same running time. 

\subsection{The k-AP Hypothesis is false}

Finally, we note that our results on hyperclique finding have applications to algorithms in additive combinatorics.

\begin{reminder}{thm:kap}
\end{reminder}

\begin{proof}
Lemma~5.7 of \cite{dudek2026classifying} states that if deciding whether a $(k-1)$-uniform hypergraph on $n$
vertices contains a $k$-hyperclique can be done in time $T(n)$, then $k$-AP Detection is in time $O(T(n^{2/k}))$. By \Cref{thm:countHC}, we have
$T(n)\leq O(n^{k-\varepsilon_k})$, so $k$-AP Detection can be solved in time
$O\bigl(n^{(2/k)(k-\varepsilon_k)}\bigr)=O(n^{2-2\varepsilon_k/k})$.
\end{proof}

\section{AI methodology and background}
On October 12 2026, we proposed using the algebraic approach to refute the hyperclique conjecture to ChatGPT 6 Pro. With a little encouragement, it obtained the main result. The full chat is available \href{https://chatgpt.com/share/6acf869d-a8f0-83ea-a00d-35492f5a3af5}{here}.

ChatGPT and Claude were both used by the authors to help understand ideas,  compare them to prior work, proofread, and draw figures. None of the text in this document is AI generated. \printbibliography

\end{document}